\documentclass[twocolumn]{aastex631}

\usepackage{newtxtext,newtxmath}
\usepackage[T1]{fontenc}
\DeclareRobustCommand{\VAN}[3]{#2}
\let\VANthebibliography\thebibliography
\def\thebibliography{\DeclareRobustCommand{\VAN}[3]{##3}\VANthebibliography}
\newcommand{\red}[1]{\textcolor{black}{#1}}

\usepackage{graphicx}	
\usepackage{amsmath}	
\usepackage{booktabs}
\graphicspath{{./}{figures/}}

\defcitealias{deWet2023}{DW23}

\received{February 18, 2025}
\revised{August 29, 2025}
\accepted{October 21, 2025}

\shorttitle{Radio afterglow of ultra-long GRB 220627A}
\shortauthors{J. K.\ Leung et al.}

\begin{document}
\title{The radio afterglow of the ultra-long GRB~220627A}
\correspondingauthor{James K.\ Leung}
\email{jamesk.leung@utoronto.ca}

\author[0000-0002-9415-3766]{James K.\ Leung}
\affiliation{David A. Dunlap Department of Astronomy and Astrophysics, University of Toronto, 50 St. George Street, Toronto, ON M5S 3H4, Canada}
\affiliation{Dunlap Institute for Astronomy and Astrophysics, University of Toronto, 50 St. George Street, Toronto, ON M5S 3H4, Canada}
\affiliation{Racah Institute of Physics, The Hebrew University of Jerusalem, Jerusalem 91904, Israel}
\affiliation{Sydney Institute for Astronomy, School of Physics, The University of Sydney, NSW 2006, Australia}
\affiliation{CSIRO Space and Astronomy, PO Box 76, Epping, NSW 1710, Australia}

\author[0000-0003-4924-7322]{Om Sharan Salafia}
\affiliation{INAF -- Osservatorio Astronomico di Brera, via E. Bianchi 46, I-23807 Merate (LC), Italy}
\affiliation{INFN -- Sezione di Milano–Bicocca, Piazza della Scienza 3, 20126 Milano (MI), Italy}

\author[0000-0002-2231-6861]{Cristiana Spingola}
\affiliation{INAF -- Istituto di Radioastronomia, via Gobetti 101, I-40129 Bologna (BO), Italy}

\author[0000-0001-5876-9259]{Giancarlo Ghirlanda}
\affiliation{INAF -- Osservatorio Astronomico di Brera, via E. Bianchi 46, I-23807 Merate (LC), Italy}
\affiliation{INFN -- Sezione di Milano–Bicocca, Piazza della Scienza 3, 20126 Milano (MI), Italy}

\author[0000-0002-2815-7291]{Stefano Giarratana}
\affiliation{INAF -- Osservatorio Astronomico di Brera, via E. Bianchi 46, I-23807 Merate (LC), Italy}

\author[0000-0002-8657-8852]{Marcello Giroletti}
\affiliation{INAF -- Istituto di Radioastronomia, via Gobetti 101, I-40129 Bologna (BO), Italy}

\author[0000-0002-8978-0626]{Cormac Reynolds}
\affiliation{CSIRO, Space \& Astronomy, PO Box 1130, Bentley, WA 6102, Australia}

\author[0000-0002-2066-9823]{Ziteng Wang}
\affiliation{International Centre for Radio Astronomy Research -- Curtin University, 1 Turner Avenue, Bentley, WA 6102, Australia}

\author[0000-0003-4341-0029]{Tao An}
\affiliation{Shanghai Astronomical Observatory, 80 Nandan Road, 200030 Shanghai, P. R. China}
\affiliation{State Key Laboratory of Radio Astronomy and Technology, A20 Datun Road, Chaoyang District, Beijing, P. R. China}

\author[0000-0001-9434-3837]{Adam Deller}
\affiliation{Centre for Astrophysics and Supercomputing, Swinburne University of Technology, John St., Hawthorn, VIC 3122, Australia}
\affiliation{ARC Centre of Excellence for Gravitational Wave Discovery (OzGrav), Hawthorn, VIC 3122, Australia}

\author[0000-0001-7081-0082]{Maria R.\ Drout}
\affiliation{David A. Dunlap Department of Astronomy and Astrophysics, University of Toronto, 50 St. George Street, Toronto, ON M5S 3H4, Canada}

\author[0000-0002-5936-1156]{Assaf Horesh}
\affiliation{Racah Institute of Physics, The Hebrew University of Jerusalem, Jerusalem 91904, Israel}

\author[0000-0001-6295-2881]{David L.\ Kaplan}
\affiliation{Department of Physics, University of Wisconsin-Milwaukee, P.O. Box 413, Milwaukee, WI 53201, USA}

\author[0000-0002-9994-1593]{Emil Lenc}
\affiliation{CSIRO Space and Astronomy, PO Box 76, Epping, NSW 1710, Australia}

\author[0000-0002-2686-438X]{Tara Murphy}
\affiliation{Sydney Institute for Astronomy, School of Physics, The University of Sydney, NSW 2006, Australia}
\affiliation{ARC Centre of Excellence for Gravitational Wave Discovery (OzGrav), Hawthorn, VIC 3122, Australia}

\author[0000-0001-5654-0266]{Miguel Perez-Torres}
\affiliation{Consejo Superior de Investigaciones Científicas (CSIC) -- Instituto de Astrofísica de Andalucía (IAA), Glorieta de la Astronomía s/n, 18008 Granada, Spain}
\affiliation{School of Sciences, European University Cyprus, Diogenes Street, Engomi, 1516 Nicosia, Cyprus}

\author[0000-0003-2705-4941]{Lauren Rhodes}
\affiliation{Department of Physics, Astrophysics, University of Oxford, Denys Wilkinson Building, Keble Road, Oxford OX1 3RH, UK}

\begin{abstract}
We present the discovery of the radio afterglow of the most distant ultra-long gamma-ray burst (GRB) detected to date, GRB~220627A at redshift $z=3.084$.  
Its prompt gamma-ray light curve shows a double-pulse profile, with the pulses separated by a period of quiescence lasting ${\sim} 15$\,min, leading to early speculation it could be a strongly gravitationally lensed GRB. 
However, our analysis of the \textit{Fermi}/GBM spectra taken during the time intervals of both pulses show clear differences in their spectral energy distributions, disfavouring the lensing scenario. 
We observed the radio afterglow from 7 to 456\,d post-burst: an initial, steep decay ($F_{\nu} \propto t^{-2}$) is followed by a shallower decline ($F_{\nu} \propto t^{-1/2}$) after ${\sim} 20$\,d. 
There are three scenarios that could explain these radio properties:   
(i) energy injection from an additional, slower ejecta component catching up to the external shock;
(ii) a stratified density profile going as $n \propto r^{-8/3}$; 
or alternatively, (iii) the presence of a slow, wide ejecta component in addition to a fast, narrow ejecta component.
We also conducted an independent test of the lensing hypothesis via Very Long Baseline Interferometry (VLBI) observations at ${\sim} 12$\,d post-burst by searching, for the first time, for multiple images of the candidate lensed GRB afterglow. 
Our experiment highlighted the growing need for developments in real-time correlation capabilities for time-critical VLBI experiments, particularly as we advance towards the SKA and ngVLA era of radio astronomy. 
\end{abstract}


\section{Introduction}
\label{sec:220627a_intro}
Gamma-ray bursts (GRBs) are produced in highly-relativistic jets launched after the core collapse of a massive star \citep{Woosley1993,Hjorth2003,Stanek2003} or the merger of binary compact objects \citep[binary neutron star or neutron star-black hole merger;][]{Eichler1989,Li1998,Abbott2017}. 
A panchromatic synchrotron afterglow \citep[e.g.,][]{Paczynski1993,Sari1998} observable from X-ray down to radio frequencies is produced when the jet decelerates into the circumburst medium, giving rise to both a reverse and forward shock at which relativistic electrons are accelerated. 
Monitoring the radio afterglow up until late times is crucial for constraining the microphysics and energetics of any given event \citep[for a review, see][]{Chandra2012}. 
The radio afterglow may also probe the reverse shock \red{\citep[e.g.,][]{Sari1999, Laskar2013, Anderson2014, vanderHorst2014, Laskar2019, Bright2023}} with greater observational latency (up to few days post-burst) compared with shorter wavelengths and allow for structure or proper motion on milli-arcsecond scales to be resolved through Very Long Baseline Interferometry (VLBI) techniques. 
However, only a small fraction of GRBs \red{${\sim}12\%$} have radio follow-up \citep[even fewer \red{${\sim}6\%$} with radio detections; \red{see the table of GRBs\footnote{\url{https://www.mpe.mpg.de/~jcg/grbgen.html}} maintained by Jochen Greiner and also}][]{Chandra2012,Anderson2018} and only a handful of GRBs have been observed with VLBI \citep[e.g., \red{GRBs 
030329A, 151027A, 170817A, 190829A, 201015A, 221009A}][]{Taylor2004,Nappo2017,Mooley2018,Ghirlanda2019,Giarratana2022,Salafia2022,Giarratana2024}. 
Now, with this paper, GRB~220627A joins this small set of events with a comprehensive radio campaign lasting more than 400\,d including high angular resolution VLBI follow-up. 

The most striking feature of GRB~220627A was the presence of two emission episodes, showing similar temporal profiles, separated by ${\sim} 15\,$min in the prompt gamma-ray light curve \citep{Roberts2022GCN32288}.
The initial trigger event on the \textit{Fermi}/Gamma-ray Burst Monitor \citep[\red{GBM, 8\,keV $-$ 40\,MeV;}][]{Meegan2009}
occurred at 21:21:00.09 UT on 2022 June 27 \citep{FGT2022GCN32278}. 
The detection of this prompt emission was later confirmed by reports of high-energy detections by the 
\textit{Fermi}/Large Area Telescope \citep[\red{LAT, operating at a higher energy range than \textit{Fermi}/GBM from $20\,$MeV $-$ $300\,$GeV;}][]{Atwood2009} \citep[][]{diLalla2022GCN32283}, 
\textit{Swift}/Burst Alert Telescope (BAT) via the Gamma-Ray Urgent Archiver for Novel Opportunities (GUANO) programme \citep{Raman2022GCN32287}, 
Konus-\textit{Wind} \citep{Frederiks2022GCN32295}, 
and the \textit{AstroSat} Cadmium-Zinc-Telluride Imager (CZTI) \citep{Suresh2022GCN32316}.
A second trigger event on the \textit{Fermi}/GBM, localised to the same region of the sky, occurred ${\sim} 15\,$min later; it was shown that this second burst episode had a similar duration\footnote{T$_{90}$ is a quantity that represents the time interval covering 90\% of the total background-subtracted counts detected by the instrument.} (T$_{90} \approx 127\,$s) and burst shape to the first episode (T$_{90} \approx 138\,$s). 
This second episode was also detected across the different energy bands of Konus-\textit{Wind}  ($20 - 100\,$keV, $100 - 400\,$keV) and \textit{AstroSat} ($20 - 200\,$keV) instruments, providing a further spectral constraint on the possible emission mechanisms. 
The two immediate competing interpretations proposed were either: (i)~a gravitationally lensed GRB event, or (ii)~an ultra-long GRB event \citep{Roberts2022GCN32288}, \red{which are conventionally characterised as those with durations greater than $1\,000$s}.

A gravitationally lensed GRB event is expected to have clear signatures imprinted on the prompt light curve: two or more segments of the light curve should show similar temporal and spectral profiles that are indicative of different arrival times for photons from different lensed images due to the different paths traversed \citep{Mao1992}. 
Given that up to 4 GRBs may be detected on any given day,
out to very high redshifts $z \sim 9$ \citep[for a brief review of high-redshift GRBs, see][]{Salvaterra2015}, a small fraction of them are expected to be gravitationally lensed: approximately 1 in 1000 GRBs \citep{Mao1992} or ${\sim} 0.3\,\textrm{sky}^{-1}\,\textrm{yr}^{-1}$ \citep{Oguri2019}. 
To date, five bursts have been claimed to be gravitationally lensed via the identification of feature similarities in different segments of their prompt light curves: GRBs 950830 \citep{Paynter2021}, 081126A \citep{Lin2022}, 090717 \citep{Kalantari2021}, 200716C \citep{Wang2021,Yang2021}, and 210812A \citep{Veres2021}. However, among them, many of these claims have been disputed on statistical grounds \citep[e.g.,][]{Mukherjee2024}. 

Lensed GRBs allow the physical properties of the intervening lens to be studied \citep[e.g.,][]{Paynter2021} and for time-delay cosmography to be performed, yielding new constraints on cosmological parameters, such as the Hubble parameter \citep{Refsdal1964}.
\red{Recent examples of time-delay cosmography have mainly been through the study of lensed supernovae --- e.g., SN Refsdal \citep{Kelly2023a,Kelly2023b}, SN H0pe \citep{Chen2024,Pierel2024a,Pascale2025}, and SN Encore \citep{Pierel2024b} --- their effectiveness for doing cosmology will improve with better precision in the time-delay measurements and improved techniques for modelling of the lenses. 
Despite the initial potential for advancing on this front,} the lensing interpretation of the GRB~220627A prompt light curve was later shown to be inconsistent with the higher-energy \textit{Fermi}/LAT data \citep{Huang2022}. 
\red{In their analysis of these \textit{Fermi}/LAT data, \citet{Huang2022}} showed that the high-energy spectrum of the burst in the first episode was inconsistent with that in the second episode: 49 photons above $100\,$MeV were detected by \textit{Fermi}/LAT in the first episode compared to none in the second.

With the lensing interpretation disfavoured, GRB~220627A was instead classified as an ultra-long GRBs, the eighth such event to be confirmed in the literature \citep[see, e.g.,][and references therein]{Campana2006,Levan2014}. 
It should be noted that GRB~220627A belongs to a less common subset of ultra-long GRBs with ``interrupted emission'' (in contrast to those with ``continuous emission''; see \citealt{Virgili2013} for a discussion): 
its prompt light curve features two pulses of activity separated by a long period of quiescence, rather than the `continuous' activity (without periods of quiescence) seen in more typical ultra-long GRB prompt light curves. 
\red{The ${\sim} 15$\,min period of quiescence between the two episodes is rather uncommon; only GRB~080407A has been reported in the literature with a longer period of quiescence (${\sim} 1\,400$\,s) between two observed emission episodes \citep{Virgili2013}.} 
It is still unclear whether the ultra-long GRB subpopulation represents the tail of the standard long GRB duration distribution \citep{Virgili2013} or a new class of GRBs with fundamentally different properties \citep{Boer2015}. 

Evidence of properties for this subpopulation differing from standard long GRBs have started to emerge.
For example, SN 2011kl, associated with ultra-long GRB~111209A, was found to be three times more luminous compared to Type Ic supernovae associated with standard long GRBs, while also possessing a very distinct supernova spectrum \citep{Greiner2015}. 
The ultra-long GRB~130925A also displayed spectral properties \citep{Bellm2014,Basak2015} seldom observed in the standard long GRB population: the X-ray spectrum transitioned from a softer spectrum (unusual for afterglows) to a harder spectrum (more typical for afterglows), which suggested the presence of either a thermal hot cocoon component \citep{Piro2014} or a dust echo scattering the prompt emission \citep{Evans2014,Zhao2014}; and the radio spectrum exhibited a sharp cutoff (decline) above 10\,GHz \citep{Horesh2015}, which 
\red{could not be explained using an electron spectral index consistent with the values inferred from the wider GRB population ($1.5 < p < 3.5$), where the electron power-law energy distribution is $dN/dE \propto E^{-p}$.} 
\red{\citet{Gendre2013} found the rate of these ultra-long events in the local Universe to be $6 \times 10^{-5}$\,Gpc$^{-3}$\,yr$^{-1}$, which was significantly lower than the local GRB rate estimates \citep[e.g., $150^{+180}_{-90}$\,Gpc$^{-3}$\,yr$^{-1}$ in][]{Howell2013}. 
The study of X-ray afterglows in the local Universe indeed found that the ultra-long events had a higher intrinsic absorption ($\textrm{NH}_{x,i} > 6\times 10^{21}\,$cm$^{-2}$) and softer photon index ($\Gamma > 3$) than the normal population; this was interpreted as indications of a turbulent mass-loss history from the progenitor leading to a dusty wind environment or simply the presence of a large stellar progenitor located in a dusty environment \citep{Margutti2015}. }

\red{More generally, the following four scenarios have been proposed to explain the long central engine activity required for ultra-long GRBs. (1) core-collapse of low-metallicity blue supergiants with much larger radii than the Wolf-Rayet progenitors of standard long GRBs: they can provide enough mass to power the central engine for thousands of seconds \citep[e.g.,][]{Gendre2013,Kashiyama2013,Nakauchi2013,Perna2018}; (2) tidal disruption of white dwarfs by black holes with mass $< 10^{5}\,M_\odot$: they can produce accretion timescales of the bound debris and energetics that are consistent with ultra-long GRBs \red{\citep[e.g.,][]{Bloom2011,Burrows2011,Zauderer2011,Levan2014,MacLeod2014,Ioka2016}}; (3)  the spin-down of a newly formed magnetar remnant following the launch of the GRB jet: the ultra-long duration of the GRB can be explained by the extra spin-down energy injection \citep{Usov1992,Greiner2015}; and (4) standard long GRB progenitor (Wolf-Rayet star) in a low-density circumburst environment (e.g., $<10^{-3}$\,cm$^{-3}$): the external shock forms further out from the progenitor in a low-density environment, allowing more time for internal shocks to dissipate radiative energy, which leads to both a higher prompt efficiency and a longer duration consistent with ultra-long GRBs \citep{Evans2014}. }

An afterglow for GRB~220627A was detected at X-ray \citep{Gropp2022GCN32296}, optical \citep{deWet2022GCN32289,NG2022GCN32304}, and radio \citep{Leung2022GCN32341,Giarratana2022GCN32454} wavelengths. 
Its localisation enabled the identification of narrow metal absorption lines within the afterglow, placing the spectroscopic redshift measurement of the afterglow at $z = 3.084$ (making it the most distant ultra-long GRB detected to date), in addition to lines from an intervening system at $z=2.665$ \citep{Izzo2022GCN32291,deWet2023}. 

In this paper, we present the results from our radio observations of GRB~220627A spanning over 400 days post-burst, which include VLBI observations intended for testing the lensing hypothesis. 
While the high-energy spectrum in this case provided sufficient data to disfavour the lensing scenario, VLBI observations offer an independent and direct test of this hypothesis, potentially free from ambiguities from light curve and spectral analysis only. 
In the case of strong lensing, these observations would enable us to model the lens, which is an important step for using lensed GRB events for time-delay cosmography. 
In Section~\ref{sec:220627a_vlbi_lensed_grb_program}, we give an overview of our ongoing VLBI program for confirming lensed GRB candidates. 
In Section~\ref{sec:220627a_observations}, we present the radio observations of GRB~220627A used in this work. 
In Section~\ref{sec:220627a_r+d}, we re-assess the \emph{Fermi}/GBM data originally used to classify GRB~220627A as a candidate lensed GRB,
discuss our physical interpretation of the radio afterglow, and described the outcomes of our VLBI experiment. 
We summarise our conclusions in Section~\ref{sec:220627a_conclusions}.

Throughout this paper, we assume a flat $\Lambda$-CDM cosmology with $H_0=~67.7$\,km\,s$^{-1}$\,Mpc$^{-1}$, $\Omega_{\text{M}} = 0.310$, and $\Omega_{\Lambda} = 0.690$ \citep{Planck2018}.

\section{Imaging Lensed GRB Afterglows with VLBI}
\label{sec:220627a_vlbi_lensed_grb_program}

GRB~220627A was the first event trigger for our ongoing VLBI programme aimed at verifying candidate lensed GRBs. 
\red{In this section, we describe why we triggered VLBI and the companion compact interferometer campaign for GRB~220627A.
In particular, we discuss how we used VLBI to test the gravitational lensing hypothesis of GRB~220627A and the broader scientific implications of confirming a lensed GRB candidate.}

\subsection{Strongly Lensed Gamma-Ray Bursts}
If a GRB is strongly gravitationally lensed, multiple images of the GRB would be produced \citep[e.g.,][]{Oguri2019}. 
In the scenario where the observing instrument does not have sufficient angular resolution to resolve the lensed images, 
the gravitational lens signature may be superimposed onto the prompt light curve \citep{Blaes1992}. 
Any flux variation from the GRB would appear at different times in the lensed images because the photons travel different path lengths. 
Therefore, in a spatially unresolved light curve there would be two pulses (for two lensed images), where the temporal profile of the second pulse should be similar to the first pulse, albeit weaker in intensity.
The spectral properties of the two pulses in this scenario should also be similar, i.e., the lens signature is consistent across all energy bands.

The time delay between the two pulses can be explained by a combination of both the Shapiro time delay and geometric effects from differing paths traversed by the two rays; it is governed by the relation provided in \citet{Krauss1991}:
\begin{equation}\label{eq:time_delay}
    \Delta t = \frac{2GM_z}{c^3}\bigg[\frac{r-1}{\sqrt{r}} + \textrm{ln}(r)\bigg]\ \textrm{s},
\end{equation}
where $M_z = M(1+z_\textrm{L})$ is the redshifted lens mass ($z_\textrm{L}$ is the lens redshift), $r$ is the brightness ratio of the two lensed images, $G$ is the gravitational constant, and $c$ is the speed of light. 

In the case of lensed GRB candidates reported to date (including GRB~220627A), the time delays between the putative lensed images are usually seconds to minutes \citep[e.g.,][]{Veres2021,Wang2021, Yang2021}. 
In this case, using Equation~\ref{eq:time_delay} and assuming a range of $r$ from 1.5 to 5 and $z$ up to ${\sim} 9$, we would infer a lens mass of ${\sim} 10^{5-7}$ M$_{\odot}$. 
We note that the optical depth of lenses in this mass range is expected to be an order of magnitude lower than those with more typical ``galactic'' masses (${\sim} 10^{12}$~M$_{\odot}$) \citep{Loudas2022}, and that none have been conclusively identified to date. 
However, this suggests that lensed GRBs may provide a promising avenue to conclusively identify the first lens within this mass range (see Subsection~\ref{subsec:broader} below).

\begin{figure}
    \centering
\includegraphics[width=0.45\textwidth,clip,trim={0.2cm 0.0cm 1.2cm 1.2cm}]{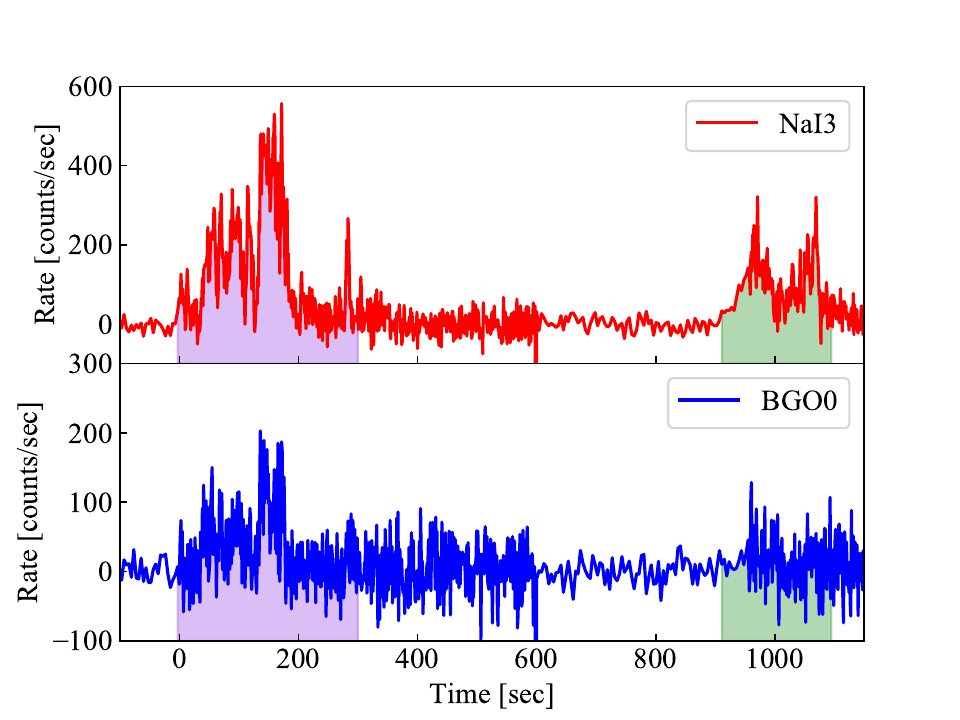}
    \caption{Background subtracted light curve of GRB 220627A from the \emph{Fermi}/GBM NaI\#3 and BGO\#0 detectors showing the count rate obtained integrating the signal in the $10-900\,$keV and $300-30\,000\,$keV energy range, respectively. 
    The shaded pink and green regions show the time intervals selected for the spectral analysis. 
    }
    \label{fig:gbm-lc}
\end{figure}

Since high-energy satellites have sub-second temporal resolutions, double-pulse prompt light curves have been the basis of various claims (and disputes) of gravitationally lensed GRBs as discussed in Section \ref{sec:220627a_intro}.
As an example, we show in Figure~\ref{fig:gbm-lc} 
the prompt light curve for GRB~220627A (the burst that is the focus of this manuscript) from \textit{Fermi}/GBM. 
The presence of two pulses with a flux ratio $r \approx 3$ \citep[this was determined by the ratio of the pulse fluences and is typical of a doubly-imaged strong lensing system; see also][]{Roberts2022GCN32288}, which showed similar temporal profiles across more than one energy band, initially provided compelling evidence supporting the strong gravitational lensing scenario.
However, such observations cannot rule out the possibility that the double-pulse profile may be produced by a physical emission mechanism intrinsic to the GRB (rather than by lensing), 
or even the possibility that the double-pulse profile does not appear in energy bands outside the observable range of the instrument (e.g., as demonstrated by GRB~220627A, the double-pulse profile was clear in the \textit{Fermi}/GBM data, but was lacking in the \textit{Fermi}/LAT data; \citealt{Huang2022}).  

The smoking gun for a gravitationally lensed GRB would be the direct detection of two (or more) consistently evolving lensed images of the afterglow with the same surface brightness.
The angular separation of the images is on the order of the Einstein radius:
\begin{equation} \label{eq:einstein_radius}
    \theta_\textrm{E} = \sqrt{\frac{4GM}{c^2}\frac{D_\textrm{LS}}{D_\textrm{L}D_\textrm{S}}} \ \textrm{rad},
\end{equation}
where $D_\textrm{LS}$, $D_\textrm{L}$ and $D_\textrm{S}$ are the distances between the lens and the GRB, the distance to the lens, and the distance to the GRB, respectively. 
For a typical event, with GRB redshift $z_\textrm{S} = 1$ and lens redshift $z_\textrm{L} = 0.4$, the Einstein radius for a $10^6\,M_\odot$ point mass lens (the typical lens mass inferred from previous claims; see above) is $\theta_\textrm{E} \approx 3\,\textrm{mas}\,(M/10^6\,M_\odot)^{1/2}(D/0.6\,\textrm{Gpc})^{-1/2}$, where $D=D_\textrm{L}D_\textrm{LS}/D_\textrm{S}$ \citep{Veres2021}. As a result, these systems are typically referred to as ``milli-lens'' systems.
Specifically for GRB~220627A, given its redshift and that of the intervening system (which we interpret here as a potential lens), the expected Einstein radius would be $\theta_\textrm{E} \approx 1.8$\,mas. 
Considering this typical angular scale expected from lensed GRB systems, direct detection of the two images in the prompt gamma-rays or through standard X-ray/optical/radio afterglow observations would be infeasible. 
VLBI is currently the only way of reaching the necessary angular resolution for confirming a strongly lensed GRB event.

To confirm a strongly lensed GRB event with VLBI observations, the detection of spatially resolved components would be required. 
Ideally, these detections are made in at least two epochs, with at least one epoch featuring quasi-simultaneous observations at two or more frequencies. 
These allow for a confirmation of lensing by checking: 

\begin{itemize}
    \item for consistency between the spectra of two lensed images \citep[e.g., figure 6 of][]{Katz1997};\footnote{In this case, we note that the time delays observed in many putative lensed GRBs -- seconds to minutes -- are much shorter than the 
    timescale on which the typical radio afterglow spectrum undergoes significant change ($\sim$days).} 
    \item for surface brightness preservation between the two lensed images (a necessary signature of gravitational lensing);
    \item that the two lensed images preserve their brightness ratio as they evolve and, aided by complementary low-resolution radio observations, evolve in accordance with known GRB closure relations \citep[e.g.,][]{Sari1998}; and
    \item that the brightness ratio of the afterglow images matches the ratio of the pulse fluences from the GRB prompt light curve (although there are certain scenarios where this may not hold, e.g., if there are significant wavelength-dependent absorption and scattering effects caused by the lens galaxy, or if the prompt emission comes from an emitting region with a large displacement from the radio emitting region -- see \citealt{Mittal2007,Cheung2014}).
\end{itemize}

The combination of all these checks makes a confirmation robust against potential ambiguities that arise from using one or two checks, such as the possibility of chance spatial association with non-afterglow related sources (e.g., a background host galaxy nucleus). 

\subsection{Broader Scientific Implications}\label{subsec:broader}

The confirmation of a strongly lensed GRB candidate has scientific implications for time-delay cosmography, the physics of the lens, and our understanding of GRBs.
A confirmation would provide a new example of gravitational lensing to test Einstein's general relativity, checking for consistency in both the temporal and spatial data.
Like other strong gravitationally lensed explosive transients, lensed GRBs have important applications in time-delay cosmography; the measured time delay can be used alongside a well-constrained Einstein radius to constrain the Hubble parameter $H_0$ \citep{Refsdal1964}. 
This is a direct consequence of the time delay between the lensed GRB images being dependent on the mass distribution of the lens, which can be modelled from the morphology and distribution of detected components in high-resolution VLBI images, as well as geometric effects from different paths in space-time, which is determined by the Hubble parameter -- for a comprehensive review of time-delay cosmography for transients, see \citet{Oguri2019}.

As GRBs are found at high redshifts up to $z \sim 9$ \citep[e.g.,][]{Salvaterra2015} and their prompt light curves (and any time delays) can be sampled accurately down to sub-second temporal resolution \citep[e.g.,][]{Meegan2009}, strongly lensed GRBs can potentially probe for lens mass and distance ranges that cannot be probed by strongly lensed quasars and supernovae (since the temporal resolution of the time delay determines the redshifted lens mass range your probe is sensitive to as shown in Equation \ref{eq:time_delay}).
Accompanying previous claims of lensed GRs were suggestions of using them to study the population statistics and physics of their lenses \citep[e.g.,][]{Paynter2021}. 
In fact, a confirmation of a lensed GRB (with an observed time delay on the order of seconds to minutes) could also lead to the first confirmation of a milli-lens object, which implies the presence of massive compact objects, such as intermediate-mass black holes (IMBH) and dark matter halos expected in some dark matter models \citep[e.g.,][]{Wilkinson2001,Spingola2019,Casadio2021}.

On the other hand, a conclusive rejection of a lensing explanation for a GRB with a double-pulse profile would put extra constraints on the physics of GRB central engines. 
For example, current GRB models that can explain multiple pulses in the prompt emission would need to be extended to explain how their emission mechanism in some GRBs can produce repeated pulses with similar temporal and spectral profiles. 
In the specific case of GRB~220627A, the comparison of the \emph{Fermi}/GBM and \emph{Fermi}/LAT data that ultimately disfavoured the lensing hypothesis \citep{Huang2022} indicated that the GRB belonged to the subpopulation of ultra-long GRBs. 

\red{Motivated by these broader scientific implications for confirming and characterising a lensed GRB event, we began our radio campaign to independently test the lensing hypothesis for GRB~220627A through VLBI shortly after the GRB was detected; this campaign is described in the following section.}

\section{Observations and Data Analysis}
\label{sec:220627a_observations}

Here, we outline our radio observations triggered for GRB~220627A as part of the VLBI programme, which aimed to provide an independent test of the lensing hypothesis to verify the conclusions from the analysis of high-energy spectra. 
The Australia Telescope Compact Array (ATCA), MeerKAT, and the Karl G. Jansky Very Large Array (VLA) observations led to the discovery of the radio afterglow and contributed to the monitoring of its evolution over a span of ${\sim}400$\,d. 
This set of observations informed our trigger of the high-resolution Very Long Baseline Array (VLBA) observations. 

\subsection{ATCA Observations}

\begin{table}
\fontsize{10pt}{12pt}\selectfont 
\caption{
Radio flux-density measurements for GRB~220627A. 
Columns 1 through 4 show the number of days \red{in the observer frame} since the \textit{Fermi}/GBM trigger (21:21:00.09 UT on 2022 June 27), the observing telescope, the central observing frequency of the radio image, and the flux-density measurements for the observation. 
We report the $3\sigma$ limit for all non-detections
\red{, where $\sigma$ here is simply the rms noise.} 
The only exception is for the VLBA observation, where a more conservative $5\sigma$ limit is provided instead.
The reported uncertainties \red{include both the statistical and systematic components added together in quadrature.} 
\red{Here, the systematic errors arising from flux-density scale calibration are typically $\lesssim 5$ per cent for the ATCA, MeerKAT, and VLA, and $\lesssim 15$ per cent for the VLBA (which is higher because it relies on system temperatures and gain curves for flux-density calibration, rather than an observation of a dedicated flux-density calibrator).}
}
\label{tab:220627a_measurements}
\begin{tabular}{cccc}
\hline\hline
$t$ (d) & Telescope &  $\nu$ (GHz) & $F_\nu$ (\textmu Jy) \\
\hline
      \phantom{0}7.5 &      ATCA &          \phantom{0}5.5 &                \red{$<98$}\phantom{0} \\
      \phantom{0}7.5 &      ATCA &          \phantom{0}9.0 &                \red{$<91$}\phantom{0} \\
      \phantom{0}7.5 &      ATCA &         16.9 &         $373 \pm \red{74}\phantom{0}$ \\
      \phantom{0}8.9 &   MeerKAT &          \phantom{0}1.3 &                \red{$<28$}\phantom{0} \\
     11.7 &      VLBA &         15.2 &               \red{$<293$} \\
     15.2 &       VLA &         10.0 &                \red{$<60$}\phantom{0} \\
     15.2 &       VLA &         15.0 &          $74 \pm \red{18}$ \\
     27.4 &      ATCA &          \phantom{0}5.5 &                $\red{<63}$\phantom{0} \\
     27.4 &      ATCA &          \phantom{0}9.0 &                $\red{<41}$\phantom{0} \\
     27.4 &      ATCA &         17.7 &          $63 \pm \red{19}$ \\
     35.7 &   MeerKAT &          \phantom{0}1.3 &                $\red{<44}$\phantom{0} \\
     39.0 &       VLA &          \phantom{0}6.0 &                $\red{<41}$\phantom{0} \\
     45.1 &       VLA &         15.1 &           $44 \pm \red{8}\phantom{0}$ \\
     455.7\phantom{0} &      ATCA &         16.9 &                $\red{<19}$\phantom{0} \\
\hline
\end{tabular}
\end{table}

We observed the field of GRB~220627A with the ATCA, using the coordinates of the reported optical afterglow \citep{deWet2022GCN32289} as the phase tracking centre.
We conducted four epochs of ATCA observations, observing the first epoch on 2022 July 5 for 10\,h \red{(7.5\,d post-burst)}, the second epoch on 2022 July 25 for 12.5\,h  \red{(27.4\,d post-burst)}, the third epoch on 2023 September 22 for 12\,h \red{(452.2\,d post-burst)}, and the fourth epoch on 2023 September 29 for 12\,h \red{(459.2\,d post-burst)}, all under project C3478 (PI: J. Leung). 
For the first two epochs, the observations were observed at multiple bands, centred on 5.5, 9, 17 and 19\,GHz, each with a bandwidth of 2\,GHz, while for the final two epochs, the observations were observed only at 17 and 19\,GHz, each with a bandwidth of 2\,GHz. 
The array was in H214 configuration for the first two observations and in H168 configuration for the final two epochs.\footnote{\url{https://www.narrabri.atnf.csiro.au/operations/array_configurations/configurations.html}} 
In these configurations, the maximum baselines are $247$ and $192\,$m, respectively. The values do not consider baselines involving antenna CA06 (all of which exceed $4\,$km) since they were flagged in our data reduction process as discussed below. 

We reduced the visibility data using standard {\sc Miriad} procedures \citep{Sault1995}. 
We used PKS B1934$-$638 to set the flux-density scale and calibrate the bandpass response, and PKS B1313$-$333 (J1316$-$3338) to calibrate the complex gains. 
Prior to imaging, we flagged out all the baselines involving antenna CA06. 
This was because the large gap in ($u,v$)-coverage between these longer baselines and the other much shorter baselines in the array created difficulties in the imaging process (due to the highly irregular shape of the resulting dirty beam). 
We produced images using the multi-frequency synthesis CLEAN algorithm \citep{Hogbom1974, Clark1980, Sault1994} and extracted flux densities in the case of a detection by fitting a point-source model in the image plane.
For the 2022 July 5 \red{(7.5\,d post-burst)}, 2023 September 22 \red{(452.2\,d post-burst)}, and 2023 September 29 \red{(459.2\,d post-burst)} observations, we did not use the 19\,GHz band as it was either severely affected by poor weather conditions or radio frequency interference. 
However, both the 17 and 19\,GHz frequency bands provided reliable data for the 2022 July 25 \red{(27.4\,d post-burst)} observations so we could image them together, providing a contiguous bandwidth of 4\,GHz centred on approximately 18\,GHz. 
In order to obtain a deep late-time image of the field, we also imaged the 2023 September 22 and 29 observations \red{(mean epoch of 455.7\,d post-burst)} at 17\,GHz together, providing an effective total integration time of 15\,h. 
The flux densities (and $3\sigma$ upper limits for non-detections) are reported in Table~\ref{tab:220627a_measurements}. 
The rms noise in all bands was higher in the first observation, compared to the second, due to poor weather conditions. 

\subsection{MeerKAT Observations}

We observed the field of GRB~220627A with the MeerKAT telescope on 2022 July 6 \red{(8.9\,d post-burst)} and 2022 August 2 \red{(35.7\,d post-burst)} under project DDT-20220705-SG-01 (PI: S. Giarratana) for 2\,h each.
These observations were centred on 1.28 GHz (L band) with a bandwidth of 0.78 GHz.
We used PKS~B1934$-$638 for bandpass and flux-density scale calibration, J1311$-$2216 for phase calibration. 
The archived raw visibilities were converted to the Measurement Set format with the KAT Data Access Library ({\sc katdal}\footnote{\url{https://github.com/ska-sa/katdal}}) for continuum imaging. 
We reduced the data using {\sc oxkat}\footnote{\url{https://github.com/IanHeywood/oxkat}} \citep[v0.3;][]{2020ascl.soft09003H}, where: the Common Astronomy Software Applications \citep[{\sc CASA};][]{McMullin2007,CASA2022} and {\sc Tricolour}\footnote{\url{https://github.com/ska-sa/tricolour}} \citep{2022ASPC..532..541H} packages were used for measurement set splitting, cross-calibration, flagging; {\sc CubiCal}\footnote{\url{https://github.com/ratt-ru/CubiCal}} \citep{2018MNRAS.478.2399K} was used for self-calibration; and {\sc WSClean} \citep{2014MNRAS.444..606O} was used for continuum imaging. All processes were executed with {\sc oxkat} L-band default settings. 
We do not detect any radio emission at the source location in both observations. 
The $3\sigma$ non-detection limits are given in Table~\ref{tab:220627a_measurements}.

\subsection{VLA Observations}

We obtained observations for GRB~220627A with the VLA at three separate epochs, with 1\,h of observation time per epoch, under project VLA/22B-031 (Legacy ID: AL1245; PI: J. Leung). 
The first epoch of observations occurred on 2022 July 13 \red{(15.2\,d post-burst)}, at the 10 and 15\,GHz (X and Ku) frequency bands, with the array in A$\rightarrow$D configuration.\footnote{This observation occurred during a reconfiguration, where the VLA antennas were transitioning from the A configuration (with a maximum baseline of 36\,km) to the D configuration (with a maximum baseline of 1\,km). For more information on VLA configurations, see: \url{https://public.nrao.edu/vla-configurations/}} 
We then obtained a second epoch at the 6\,GHz (C band) frequency band on 2022 August 5 \red{(39.0\,d post-burst)} and a third epoch at the 15\,GHz (Ku band) frequency band on 2022 August 11 \red{(45.1\,d post-burst)}; the array was in D configuration for both the second and third epoch of observations. 
For each observation, we used 3C 286 to calibrate the bandpass response and flux-density scale; we also used PKS B1313$-$333 (J1316$-$3338) to calibrate the complex gains.

Prior to imaging, we obtained the pipeline-calibrated\footnote{All VLA data is calibrated by an automated {\sc CASA} pipeline \citep{Kent2020}. For more information, see: \url{https://science.nrao.edu/facilities/vla/data-processing/pipeline}} visibility data from the data archives (see Data Availability section). 
We used {\sc CASA} to verify the quality of the pipeline-calibrated data and also to image the field using the wide-field, multi-frequency synthesis CLEAN algorithm. 
For the 2022 July 13 \red{(15.2\,d post-burst)} observation, due to problems associated with the unbalanced $(u,v)$-coverage of the hybrid array, we were only able to obtain reliable data by using just the antennas that had already moved to the D configuration (i.e., we flagged all antennas located more than 2\,km from the centre of the array). 
After further accounting for antennas under maintenance and antennas flagged by the automated pipeline, the imaging process for the 10 and 15\,GHz observations on 2022 July 13 \red{(15.2\,d post-burst)} used data from only 12 and 13 antennas, respectively; for the 2022 August 5 \red{(39.0\,d post-burst)} observation (at 6\,GHz), the imaging process used data from 21 antennas; and for the 2022 August 11 observation \red{(45.1\,d post-burst)} at 15\,GHz, the imaging process used data from 22 antennas. 
Further self-calibration was not performed for all observations since weak field sources in addition to the presence of poorly-modelled extended sources led to limited improvements in the gain solutions. 
Due to the presence of imaging artefacts caused by the sidelobes of a bright, extended source outside the primary beam, PKS~J1325$-$3236, we phase rotated the 2022 August 5 (6-GHz band) visibility data to J2000~13:25:32.8,$-$32:30:45 before imaging so we could deconvolve both the target and PKS~J1325$-$3236 in our CLEAN cycles. 
For each observation, we imaged using the full bandwidth; i.e., 4\,GHz for the observations centred on 6 and 10\,GHz and 6\,GHz for the observations centred on 15\,GHz. 
In the case of a detection, we extracted the flux densities by fitting a point-source model in the image plane and report them in Table~\ref{tab:220627a_measurements} (for non-detections, we give the $3\sigma$ upper limit). 

\subsection{VLBA Observations}
\label{ssec:220627a_vlba}

We obtained high-resolution observations for GRB~220627A with the VLBA on two consecutive days under project VLBA/22B-032 (Legacy ID: BL296; PI: J. Leung). 
These observations were 6\,h in duration, starting at 22:30 UT on both 2022 July 8 and 9 \red{(corresponding to a mean epoch of 11.7\,d post-burst)}. 
All standard VLBA antennas,\footnote{The standard array of antennas consists of 10 stations, see: \url{https://science.nrao.edu/facilities/vlba/docs/manuals/oss/sites}} 
with the exception of the Kitt Peak station which was out due to wildfires, were available for both observations. 
The longest baseline was from Mauna Kea to Saint Croix, giving a projected baseline of ${\sim} 8\,600$\,km or ${\sim} 430$\,M$\lambda$. 
For these observations, the VLBA was configured to use the digital downconverter (DDC) observing system, observing at the central frequency of 15.168\,GHz (Ku band), with a total bandwidth of 512\,MHz (split into four 128\,MHz sub-bands) in two circular polarisations (right- and left-circular polarisations). 
Using a 2-bit sampler, the resulting data rate was 4\,Gbps. 
The data were correlated after the observations using DiFX \citep{Deller2007} with an integration time of 1\,s and a spectral resolution of 1\,MHz at Socorro, New Mexico, USA. 

We carried out standard reduction procedures on the correlated datasets from both observations simultaneously using {\sc AIPS} \citep{Greisen2003}. 
We applied standard \textit{a-priori} amplitude calibration using the provided system temperature and gain curve tables. 
The instrumental delays were determined using the fringe finders J0555$+$3948 and J1642$+$3948. 
The latter was also used for solving the bandpass. 
Since the expected flux density of our target, GRB~220627A, was $< 0.5\,$mJy as informed by our earlier ATCA observations, we phase-referenced our experiment using the calibrator J1324$-$3235, which was observed in 45\,s scans as part of 100\,s duty cycle. 
Although this calibrator had a low unresolved flux-density measurement at 8.6\,GHz (50\,mJy), it was chosen due to its close separation ($0.32\degr$) from the target position, which was particularly important given the low elevation of the source with respect to the VLBA antenna stations. 
To attain better signal-to-noise, we combined all the sub-bands during our global fringe fitting process on the fringe finders and phase-reference calibrators in order to solve for the residual phases, delays, and rates. 
The solutions from the phase-reference calibrator were interpolated to both the target and our check source, PKS B1313$-$333 (J1316$-$3338), located $1.75\degr$ away from the target. 
Finally, we used {\sc Difmap} \citep{Shepherd1997} with a natural weighting scheme (after trying several other robust weighting schemes to search for sources in the field of interest) to obtain the final images of both the target and the check source. 
The resulting image had a beam size of $0.49\,\textrm{mas} \times 1.51\,\textrm{mas}$ with a position angle of $-5.4\degr$.

In our VLBA observations, we were unable to identify any sources that were related to the GRB~220627A event within the optical\footnote{Optical position: $\alpha = 13$:25:28.49, $\delta = -32$:25:33.31, uncertainty of 0.1$\arcsec$ for each coordinate \citep[][]{deWet2022GCN32289}.} and radio\footnote{Radio position: $\alpha = 13$:25:28.57, $\delta = -32$:25:33.31, uncertainty of 0.1$\arcsec$ on $\alpha$ and 0.2$\arcsec$ on $\delta$ (from our highest signal-to-noise VLA detection).} positional uncertainties. 
The final target image achieved an rms sensitivity of $51\,$\textmu Jy\,beam$^{-1}$ and no noise peaks exceeded the $5\sigma$ level (i.e., ${\sim} 250$\,\textmu Jy\,beam$^{-1}$). 
After performing self-calibration iterations on the check source (located further from the phase-reference calibrator than the target), we found that the phase errors were no more than ${\sim} 10\degr$ for most antennas, suggesting that possible problems with phase transfer were not responsible for the non-detection.
The $5\sigma$ upper limit is reported in Table~\ref{tab:220627a_measurements}. 
We verified this result in an independent calibration of the data (as a blind test) using standard reduction procedures in {\sc CASA}; we attained a similar rms sensitivity (and upper limit), arriving at the same conclusions. 

\section{Results and Discussion}
\label{sec:220627a_r+d}

\subsection{\fontsize{9.5}{12}\selectfont Re-analysis of \textit{Fermi}/GBM Observations and Classification as an Ultra-long GRB}
\label{ssec:220627a_fermigbm}

\begin{table*}
\begin{tabular}{ccccccc}
\hline
Interval & log(L) & $E_{\rm peak}$ & $\alpha$ & $\beta$ & factor & PG-stat(dof) \\
(sec) & (erg/s) & (keV) & & & \\
\hline
$[-6, 300]$ & $52.37^{+0.04}_{-0.04}$ & $282.8^{+33.1}_{-29.6}$ & $0.95^{+0.06}_{-0.06}$ & $2.7\red{4}^{+0.39}_{-0.28}$ & $1.10^{+0.16}_{-0.13}$ & 704(219) \\
$[910, 1096]$ & $52.17^{+0.09}_{-0.09}$ & $366.1^{+803.6}_{-143.6}$ & $1.31^{+0.10}_{-0.14}$ & $2.56^{+2.03}_{-0.51}$ & $0.67^{+0.31}_{-0.22}$ & 337(219) \\
\addlinespace[0.5mm]
\hline
\end{tabular}
\caption{Spectral parameters of the fit of the prompt emission spectrum of the two emission episodes. Errors represent the 95\% confidence interval for each parameter value.}
\label{tab:gbm}
\end{table*}

\begin{figure*}
    \includegraphics[width=\textwidth]{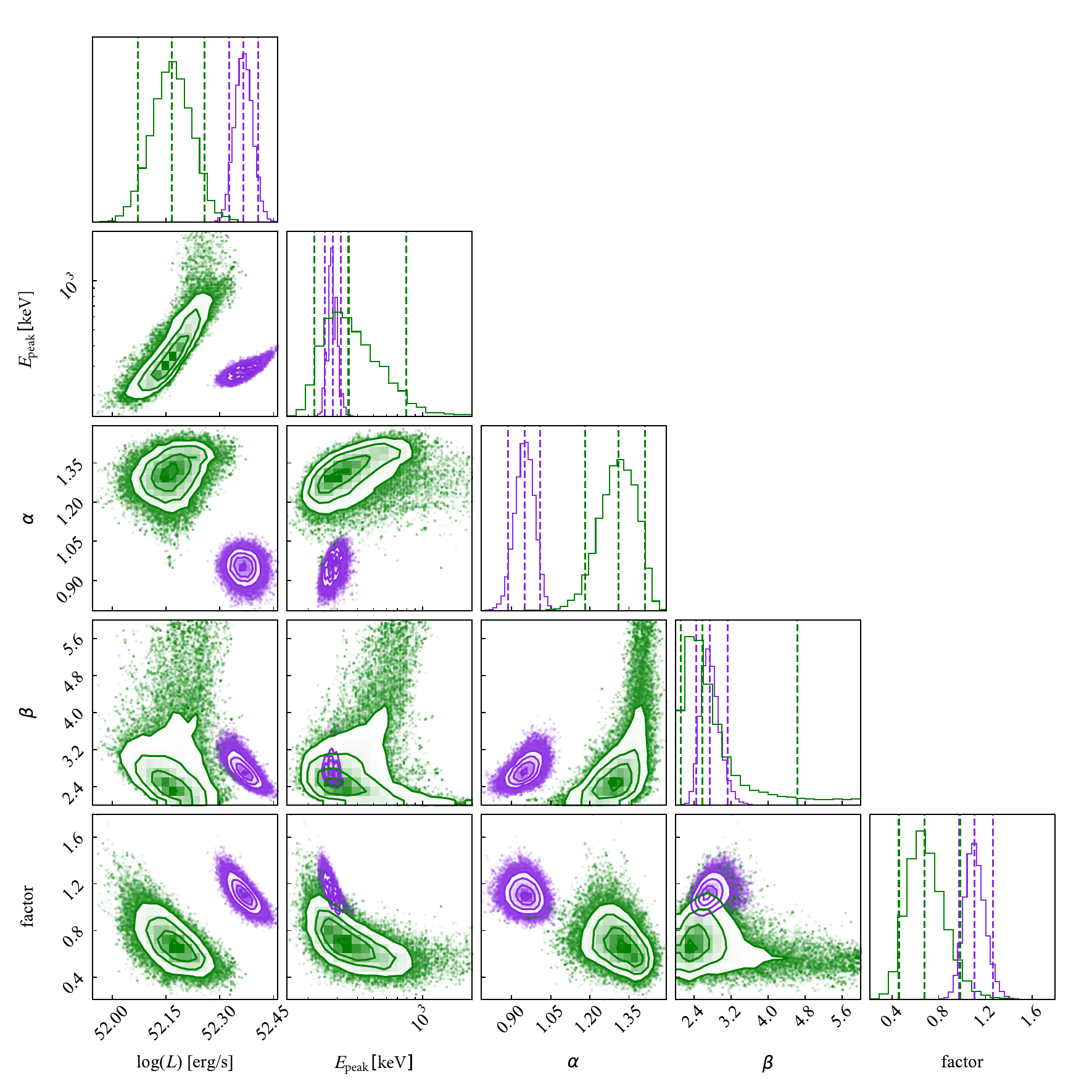}
\caption{Posterior distributions of the parameters of the spectral fit to the two time intervals (violet and green for interval 1 and 2 respectively, see Table~\ref{tab:gbm}). The median and 95\% confidence intervals are marked by the dashed vertical lines in the histograms.}
\label{fig:gbm-cont}
\end{figure*}

\begin{figure}
    \centering
    \includegraphics[width=0.45\textwidth,clip,trim={0.4cm 0.0cm 1.25cm 1.1cm}]{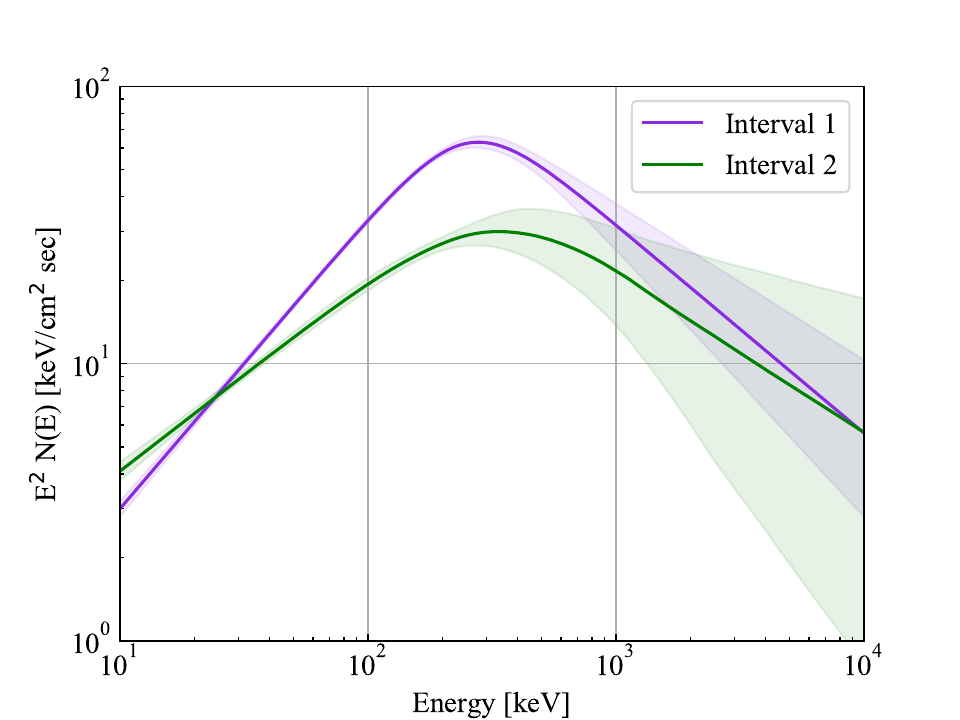}
    \caption{Spectra energy distribution of the prompt emission spectrum. The two time intervals analysed are shown. The shaded regions represent the 68\% confidence interval on the best fit model (solid lines).}
    \label{fig:gbm-sed}
\end{figure}

GRB~220627A was originally proposed as a candidate lensed GRB based on similar pulse profiles observed in the \emph{Fermi}/GBM data \citep{Roberts2022GCN32288}. 
While this interpretation was later disfavoured after assessing the spectra inferred for both pulses using data from \emph{Fermi}/LAT \citep{Huang2022}, we check to see whether a similar conclusion could have been drawn from the \emph{Fermi}/GBM data alone.

In particular, we analyse the \textit{Fermi}/GBM data of both the NaI\#3 and BGO\#0 detectors. 
\red{We used the \textit{Fermi} \textsc{Science Tools} v.2.2.0 to analyse the \textit{Fermi}/GBM data following the standard procedure\footnote{\url{https://fermi.gsfc.nasa.gov/ssc/data/analysis/software/}} \citep[e.g.,][]{Ravasio2019,Ravasio2024}. We extracted the light curves of NaI\#3 in the $10-900$\,keV energy range and of BG\#0 in the $300-30\,000$\,keV energy range. }
The background subtracted light curve of these two detectors is shown in the top and bottom panels, respectively, of Figure~\ref{fig:gbm-lc}. 
The prompt emission shows two distinct emission episodes, each lasting few hundred seconds, separated by a much longer period of quiescence. 
The two emission episodes have quite different temporal structures, with the first one being more variable and the second one apparently composed by two sub-pulses. 

We extracted the time integrated spectrum of the two emission episodes by selecting two time intervals, [$-6$,300] and [910,1096] seconds, shown by the shaded regions in Figure~\ref{fig:gbm-lc}. 
Given the large separation of the two time intervals, the background spectrum was estimated by considering four intervals bracketing the two emission episodes: [$-450$,$-200$], [400,500], [700,850], and [1200,1300] seconds. 
A third order polynomial was fit to the spectra contained in these time intervals and then interpolated to the time region selected for the spectral analysis. 

We performed the spectral analysis with \textsc{Xspec} v.12.14.0 \citep{Arnaud1996}. We considered the energy range between 10 and 900 keV for the NaI spectrum and excluded the $30$ to $40\,$keV range due to the contamination by the Iodine K-edge at such energies. The energy range 300\,keV to 30\,MeV was considered for the BGO spectrum. We allowed for a free inter-calibration constant factor between the NaI and BGO  spectra. 
We adopted the PG-stat, i.e., Poisson statistic for the source counts and Gaussian statistic for the background counts, in performing the fit. 

The spectral model adopted is a smoothly broken power law: 
\begin{equation}
\frac{dN}{dE} = A E_{\rm break}^{-\alpha}\left[ \left(\frac{E}{E_{\rm break}}\right)^{\alpha n}+\left(\frac{E}{E_{\rm break}}\right)^{\beta n}\right]^{-\frac{1}{n}}
\label{sbpl}
\end{equation}
in photons cm$^{-2}$ s$^{-1}$ keV$^{-1}$ units. $\alpha$ and $\beta$ are the spectral index of the power laws below and above, respectively, the characteristic break energy $E_{\rm break}$. The model parameter $n$ sets the curvature between the two power laws. We assume $n=2$ which resembles the curvature of the typical Band function which has been fitted to GRB spectra (\citealt{Band1993}, see also \citealt{Ravasio2018}).

The normalization constant in Equation~\ref{sbpl} is defined as: 
\begin{equation}
    A={{L}\over{4\pi d_{L}(z)^2\int_{1\,{\rm keV}/(1+z)}^{10\,{\rm MeV}/(1+z)} E\times{{dN}\over{dE}}{dE}}}
\end{equation}
where $L$ is the luminosity in units of [erg/s] in the 1\,keV$-$10\,MeV rest frame energy range and $d_{L}(z)$ is the luminosity distance corresponding to the redshift $z$ of the source. Therefore, the luminosity $L$ is one of the four free parameters of the spectral model. 

The peak energy of the spectral energy distribution (SED), i.e., the energy where $E^2\times dN/dE$ peaks, is: 
\begin{equation}
    E_{\rm peak}={{E_{\rm break}}\over{\left({{\alpha-2}\over{2-\beta}}\right)^{{1}\over{n(\alpha-\beta)}}}}
\end{equation}

We report the fit results for the two time intervals in Table~\ref{tab:gbm}. 
The posterior distributions of the fit parameters are shown in Figure~\ref{fig:gbm-cont} and the resulting fits are shown in Figure~\ref{fig:gbm-sed}. 
In addition to the aforementioned differences in temporal structures between the first and second episodes (as shown in Figure~\ref{fig:gbm-lc}), the spectra taken at the two time intervals are clearly distinct from each other, e.g., the $\alpha$ slope parameter differs between the two intervals with a significance $>5\sigma$. 
It is likely that the initial lensing hypothesis proposed in \citet{Roberts2022GCN32288} was based off the analysis of the \textit{Fermi}/GBM data with lower-resolution binning, where the temporal structures between the two episodes are more similar. 
Therefore, based on the re-analysis of the \textit{Fermi}/GBM data alone -- without requiring the additional information from \textit{Fermi}/LAT as in \citet{Huang2022} -- we can reject the lensed GRB hypothesis and verify the ultra-long GRB classification.

\subsection{\fontsize{9.5}{12}\selectfont Phenomenology and Interpretation of the Radio Afterglow}
\label{ssec:radio-ag}

In addition to informing our VLBI observations, our broadband radio light curves provide a detailed view of the afterglow of GRB~220627A between $\sim$4 and 450 days post-burst. 
Here we assess implications of these observations on the properties of the afterglow and the progenitors of ultra-long GRBs.

\begin{figure*}
    \includegraphics[width=\textwidth]{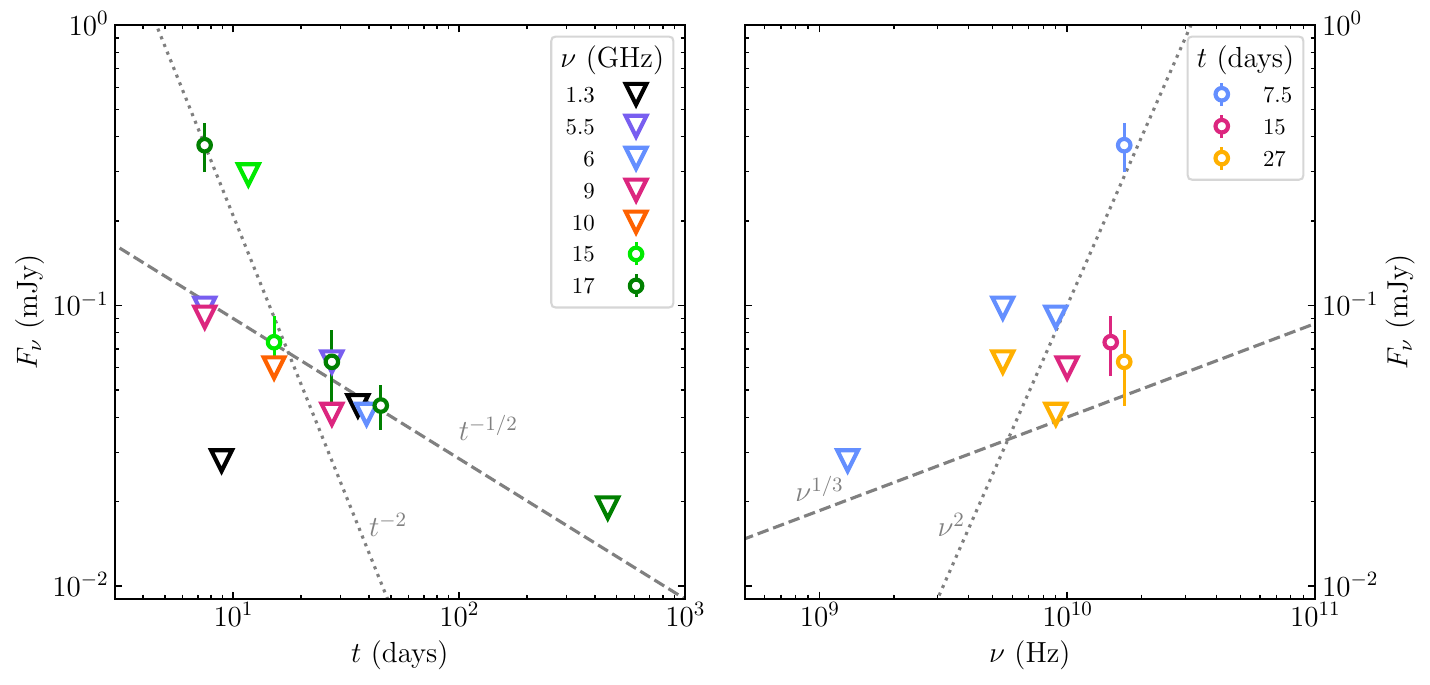}
    \caption{
    \textbf{\textit{Left}:} Radio light curves for GRB~220627A. 
    Radio detections are represented with circular markers. $3\sigma$ upper limits, represented using downward-pointing triangular markers, are used for all non-detections, with the exception of the VLBA observation ($t=11.7$ d, $\nu=15.2$ GHz), where a $5\sigma$ upper limit is used instead. 
    The grey dotted and dashed lines show a $t^{-2}$ and a $t^{-1/2}$ slope, respectively, for reference.
    \textbf{\textit{Right:}} Radio spectra for GRB~220627A.
    The plot shows different spectral snapshots centred around 3 different times as indicated in the legend.  
    Markers carry the same meaning as those in the light curves plot. 
    The grey dotted and dashed lines show a $\nu^{2}$ and a $\nu^{1/3}$ slope, respectively, for reference.
    }
    \label{fig:220627a-radio-only}
\end{figure*}

\subsubsection{Basic Properties of the Radio Afterglow}

Radio frequency light curves and spectra of the GRB~220627A afterglow at 7.5, 15, and 27\,d post-burst are shown in Figure~\ref{fig:220627a-radio-only}. 
Our ATCA observations 7.5\, post-trigger (light blue spectrum in the \textit{right} panel of Figure~\ref{fig:220627a-radio-only}) indicate a positive spectral slope $\alpha>1.5$ between 9 and 17 GHz, where $F_\nu\propto \nu^\alpha$ and we quote a 3-sigma lower limit. 
Assuming synchrotron emission, this is a strong indication of self-absorption at these frequencies and at this epoch, with the self-absorption frequency $\nu_\mathrm{a}> 17$ GHz. At later times, our observations still indicate a rising spectrum, but the available limits can also accommodate a shallower slope of $F_{\nu} \propto\nu^{1/3}$ (Figure~\ref{fig:220627a-radio-only}, \textit{right}). Due to a lack of detections, we cannot constrain the temporal evolution at $\nu<15$\,GHz. The $15-17$\,GHz light curves, on the other hand, reveal an unexpected flattening in the decay of the radio flux density at $t\gtrsim 15$\,d (Figure~\ref{fig:220627a-radio-only}, \textit{left}): while the evolution between 
7.5 and 15\,d is consistent with a steep $F_{\nu} \propto$ $t^{-2}$ decay, the subsequent detections indicate a shallower $F_{\nu} \propto$ $t^{-1/2}$. 

\subsubsection{Comparison to the Afterglow Model of \citet{deWet2023}}

Previously, \citet[][\citetalias{deWet2023} hereafter]{deWet2023} presented a best-fit model for the afterglow of GRB~220627A. 
Specifically, they used the \textsc{ScaleFit} software package \citep{Ryan2015} to model a set of X-ray, optical, and radio afterglow data (their radio data points are a preliminary subset of what we present in this paper). 
Overall, their modelling favoured a homogeneous circumburst medium and converged to the following afterglow model parameters: 
slope of the electron energy distribution $p = 2.00^{+0.08}_{-0.06}$,
the isotropic-equivalent kinetic energy of the outflow $E_\textrm{iso} = 9^{+65.1}_{-7.4} \times 10^{53}$\,erg,
circumburst medium density $n_0 = 19^{+133}_{-16}$\,cm$^{-3}$               
the microphysical shock parameters relating to the fractional internal energy in accelerated electrons and magnetic fields $\bar\epsilon_\textrm{e}~=~\frac{p-2}{p-1} \epsilon_\textrm{e}~=~1.5^{+8.1}_{-1.3}\times 10^{-3}$ and $\epsilon_\mathrm{B}=4.6^{+30.7}_{-4.3}\times 10^{-3}$,
jet half-opening angle $\theta_\textrm{j} = 4.5^{+1.2}_{-0.3}\degr$,
and host galaxy extinction $A_\textrm{V,host} = 0.09  \pm 0.03$\,mag.

\begin{figure*}
    \includegraphics[width=\textwidth,clip,trim={0mm 2mm 0mm 0mm}]{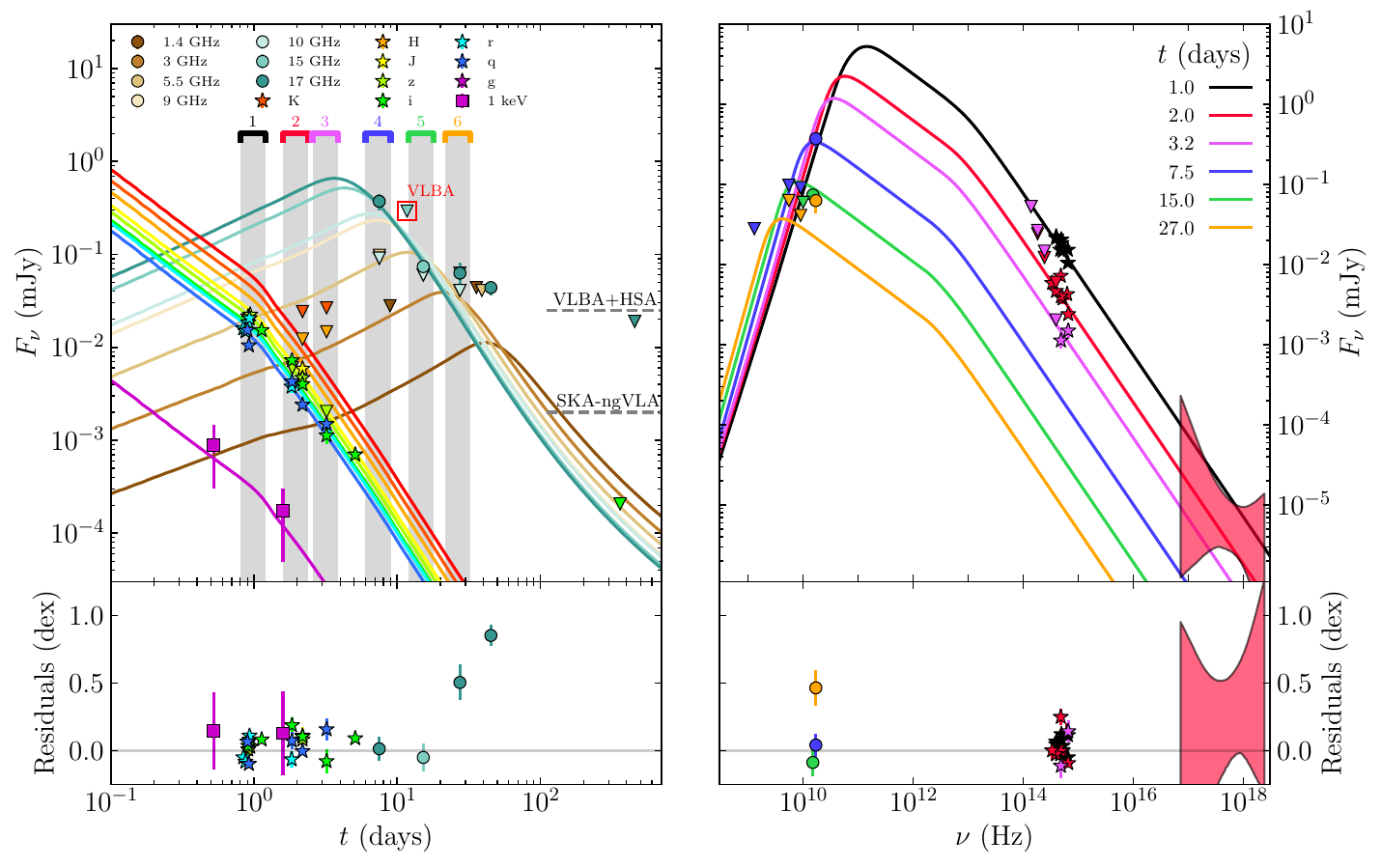}
    \caption{
    \textbf{\textit{Left}:} Multi-wavelength light curves for GRB~220627A and forward shock synchrotron model from \citetalias{deWet2023}. 
    The radio data points are represented with circular markers, the near-infrared and optical data points with star markers, and the X-ray data points with square markers. 
    $3\sigma$ upper limits, represented using downward-pointing triangular markers, are used for all non-detections, with the exception of the VLBA observation (highlighted with a red square), where a $5\sigma$ upper limit is used instead. 
    The dashed lines represent the $5\sigma$ sensitivity thresholds expected for the VLBA+HSA and SKA-VLBI/ngVLA-LBA arrays (in 10 and 1\,h of integration, respectively).
    \textbf{\textit{Right:}} Multi-wavelength SED snapshots for GRB~220627A, from radio frequencies through to X-ray energies.
    The plot shows different spectral snapshots of the afterglow centred around 6 different times as indicated in the legend. 
    The temporal window for data points included in each SED snapshot is indicated by the shaded vertical bands in the light curve plot. 
    Markers in the SED snapshots plot carry the same meaning as those in the light curves plot. 
    \red{The residual plots on the bottom, showing the difference between the data and the model for all detections in logarithmic space, highlight the discrepancy between the shallow radio evolution after 15\,d post-burst and the synchrotron model from \citetalias{deWet2023}.}
    }
    \label{fig:220627a-lcsed}
\end{figure*}

In Figure \ref{fig:220627a-lcsed}, we plot the radio data from Table~\ref{tab:220627a_measurements}, optical data from \citetalias{deWet2023}, and the X-ray data from the \textit{Swift} Burst Analyser\footnote{\url{https://www.swift.ac.uk/burst_analyser/00021506/}} \citep{Evans2010}. 
Also shown are model light curves and spectra obtained with \textsc{BOXFIT}\footnote{The use of \textsc{BOXFIT} should ensure that the predictions closely match those from \textsc{ScaleFit} (i.e., the code employed by \citetalias{deWet2023}), as the latter is based on the former.} v.\ 2.0 \citep{vanEerten2012}, with the same parameters as the best-fit model of \citetalias{deWet2023}.

Our 7.5\,d radio spectrum (epoch 4 in Figure~\ref{fig:220627a-lcsed}) is in tension with the predictions from the \citetalias{deWet2023} model, which over-predicts the flux density at $10$\,GHz by at least a factor of ${\sim}2$ at this time. 
Moreover, the temporal evolution revealed by our $15-17$\,GHz observations does not match that predicted by the model: the relatively early jet break in the \citetalias{deWet2023} model causes the radio light curve to drop steeply as $F_\nu\propto t^{-p}$ (where $p=2.0$ in their model) after ${\sim}5-7$ days (corresponding to the time in their model after which both the injection frequency, $\nu_\mathrm{m}$, and the self-absorption frequency, $\nu_\mathrm{a}$, cross the observing band). 
While this is consistent with the observed decay between 7.5 and 15 days, the subsequent flatter evolution remains unexplained. 
We note that we found a similar disagreement also with the alternative model presented in appendix A of \citetalias{deWet2023}, which assumes a wind-like external medium. 

Given that the flux density at 17\,GHz is seen to vary, with a decay at least as steep as $F_{\nu} \propto t^{-1/2}$ from 27 to 445\,d (see Figure~\ref{fig:220627a-radio-only}, \textit{left}), and given the positive spectral slope implied by our upper limits at lower frequency, we can exclude that the excess emission with respect to the \citetalias{deWet2023} model is due to star formation in the host galaxy.

\begin{figure*}
    \centering
    \includegraphics[width=\textwidth,clip,trim={0mm 2mm 0mm 0mm}]{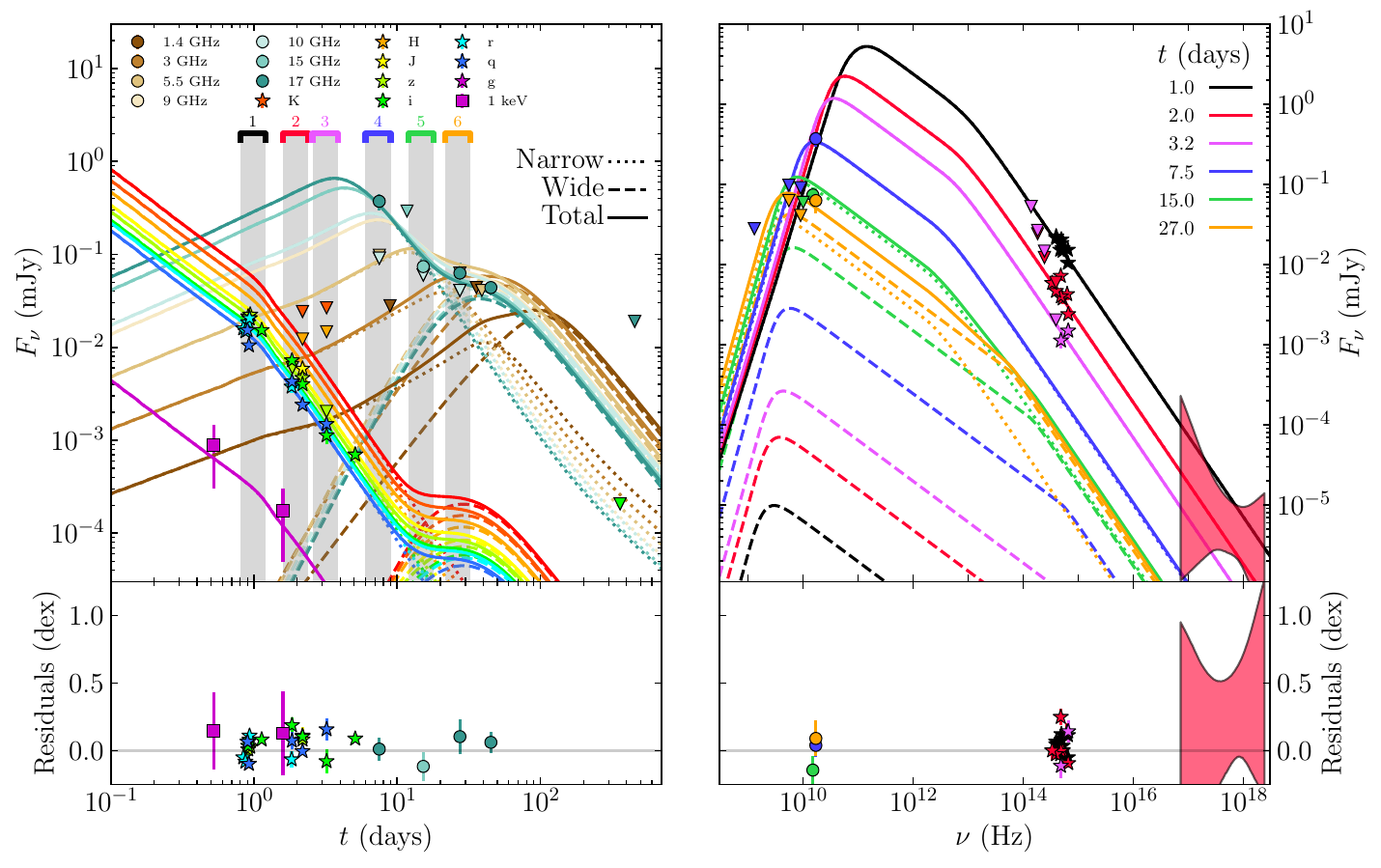}
    \caption{Model with additional wider jet component \red{better explains the shallow radio evolution observed in the afterglow after 15\,d post-burst}. The fast, narrow jet model of \citetalias{deWet2023} features an isotropic-equivalent energy of $E_\mathrm{iso}=9\times 10^{53}$ erg and a half-opening angle $\theta_\mathrm{j}=4.5\degr$. The additional wide jet component has $E_\mathrm{iso,w}=3\times 10^{53}$ erg, $\theta_\mathrm{j,w}=15\degr$, and an initial bulk Lorentz factor $\Gamma_\mathrm{0,w}=3$. The shock microphysical parameters are the same for the two components, namely $\epsilon_\mathrm{e}=0.076$, $\epsilon_\mathrm{B}=4.6\times 10^{-3}$ and $p=2.02$.}
    \label{fig:narrow+wide}
\end{figure*}

In addition, at least two of our upper limits below $17$\,GHz (at 7.5 and 15\,d) are inconsistent with the \citetalias{deWet2023} model. However, given the good match between the model and higher-frequency observations at these epochs, a possible explanation of such inconsistency could be traced back to the simplified description of the electron population in the afterglow model: the energy distribution of electrons in GRB afterglow models \citep[e.g.]{Sari1998,Panaitescu2000,vanEerten2012} is described as a non-thermal power law, but physically we expect a lower-energy thermal population to be present in these relativistic outflows \red{\citep[e.g.,][]{Giannios2009,Sironi2013,Warren2017,Gao2024,Rhodes2025}}. Even in the case in which the latter electrons constitute a non-dominant fraction over the total, with a negligible contribution to the radiated power, they could still affect the low-energy spectrum by increasing the self-absorption optical depth \red{\citep[][]{Eichler2005,Ressler2017,Warren2018,Warren2022,Margalit2021,Margalit2024}}. For that reason, we do not consider these inconsistencies as severe.

\subsubsection{\red{Explaining the Shallow Radio Decay}}

\red{We discuss below three scenarios that can explain the shallow decay observed in radio at $t>15\,\mathrm{d}$. For each scenario, we describe the modelling choices and the implications on our understanding of the source.}

\paragraph{\red{Energy injection from an additional, slower ejecta component}}
\red{the decay of the radio light curve could become shallower due to the energy injection occurring when a slower ejecta component catches up to the external shock \citep{Panaitescu1998,Sari2000}. First, we can estimate the Lorentz factor of such ejecta for the collision to happen at an observed time $t_\mathrm{coll}\sim 15\,\mathrm{d}$, assuming a similar ejection time as the main ejecta that caused the peak of the afterglow emission. Assuming the dynamics of the blast wave to be described, before the collision, by the \citet{Blandford1976} self-similar solution in a homogeneous external medium, at an observer time $t$ the part of the blast wave closest to the observer is located \citep{Sari1998} at a radius $r(t)\sim (17 E_\mathrm{iso} t /(1+z) 4\pi m_\mathrm{p} c n_0)^{1/4}$ from the central engine, where $m_\mathrm{p}$ here is the proton mass. The radius of the slower ejecta component is instead $r_\mathrm{sl-ej}(t)\sim \Gamma_\mathrm{sl-ej}^2 c t/(1+z)$. Equating these two radii and setting $t=t_\mathrm{coll}$, we obtain the required ejecta Lorentz factor:}
\begin{align}
    \Gamma_\mathrm{sl-ej} &\sim \left(\frac{17(1+z)^3 E_\mathrm{iso}}{4\pi m_\mathrm{p}c^5 n_0 t_\mathrm{coll}^3}\right)^{1/8}\\
    & \approx 10\,E_\mathrm{iso,54}^{1/8}n_0^{-1/8}\left(\frac{1+z}{4.084}\right)^{3/8}\left(\frac{t_\mathrm{coll}}{15\,\mathrm{d}}\right)^{-3/8}. \nonumber
\end{align}
\red{In words, ejecta as fast as $\Gamma_\mathrm{sl-ej}\sim 10$ catch up with the blast wave as late as $t\sim 15\,\mathrm{d}$. We describe the energy injection as an increase in the isotropic-equivalent blast wave energy with observer time}\footnote{\red{The reason why it makes sense to parametrize the blast wave energy as a function of the observer-frame time is because the arrival time of a photon emitted at a lab-frame time $t_\mathrm{lab}$ and at a typical latitude $\theta=1/\Gamma$, $t=t_\mathrm{lab}-\beta r/c$, is almost identical to the `retarded' time $t_\mathrm{r}=t_\mathrm{lab}-\beta_\mathrm{ej}r/c$ that represents the ejection time of relativistic ejecta travelling at $\beta_\mathrm{ej}$ that catch up with the blast wave at a radius $r$.}} \red{at $t>t_\mathrm{coll}$, parametrized as $E_\mathrm{iso}(t)=E_\mathrm{iso}\times(t/t_\mathrm{coll})^\delta$. This modifies the blast wave Lorentz factor evolution such that the shock downstream Lorentz factor scales as $\Gamma(t)\propto t^{(4\delta-3)/8}$ and the blast wave radius as $r\propto t^{(\delta+1)/4}$. In the post-jet-break phase, neglecting sideways expansion of the shock and focussing on the $\nu_\mathrm{a},\nu_\mathrm{m}< \nu < \nu_\mathrm{c}$ spectral regime, the flux density is proportional to $F_\nu \propto \Gamma^{2(p+1)}r^{3}$, which leads to a light curve scaling $F_\nu \propto t^{((7+4p)\delta-3p)/4}$. With $p\sim 2$, a scaling $F_\nu\propto t^{-1/2}$ can be reproduced by setting $\delta \sim 4/15 \approx 0.27$. This scaling is expected \citep{Panaitescu1998,Sari2000,Zhang2006} if the slow ejecta features a velocity profile such that their mass above a Lorentz factor $\Gamma$ is $M(>\Gamma)\propto \Gamma^{-s}$ with $s=(7\delta+3)/(3-\delta)\sim 73/41 \approx 1.8$. Since the observer time increases fourfold during the shallow decay phase, the energy increase due to injection is just $\sim 4^{4/15} \approx 145\%$.}

\red{These calculations show that a relatively slow shell ejected shortly after the main ejecta and containing about half as much energy, could produce the observed flattening in the radio light curve. The presence of a tail of slower ejecta carrying a substantial energy is a common expectation from the unsteady central engines behind leading GRB jet prompt emission scenarios, such as the internal shocks scenario \citep{Rees1994} and the internal-collision-induced magnetic reconnection and turbulence \citep[ICMART;][]{Zhang2011} scenario. 
Unfortunately, model predictions are not yet sufficiently detailed as to answer the question whether the requirements found here are easily satisfied in a typical central engine.}

\paragraph{\red{Changing external density profile}}
\red{the flattening in the radio light curve decay could stem from a steepening in the external density profile, which could signal, for example, the fact that the blast wave is emerging from the molecular cloud where the progenitor has formed. Assuming the external density to scale as $n\propto r^{-k}$, the Lorentz factor in the shock downstream then evolves \citep{Blandford1976} as $\Gamma\propto r^{-(3-k)/2}$. In terms of observer frame time, we then have $r\propto t^{1/(4-k)}$ and $\Gamma\propto t^{-(3-k)/2(4+k)}$. In the post-jet-break phase the flux density evolves as $F_\nu\propto \Gamma^{2(p+1)}r^{3-k}$ (the $3-k$ exponent here reflects the slower increase in the number of shocked electrons with radius when $k>0$), so that $F_\nu \propto t^{-(3-k)p/(4-k)}$. This shows that the $t^{-1/2}$ decay can be reproduced, assuming $p\sim 2$,  if the external density decreases with a slope $k = (6p-4)/(2p-1)\sim 8/3$. }

\paragraph{\red{An additional, wider ejecta component}}
\red{as an alternative scenario, we consider the possibility of a} wider and slower component in the ejecta, on top of the fast and narrow component whose afterglow emission is described by the \citetalias{deWet2023} model. These ejecta decelerate at a larger radius with respect to the narrow jet, enhancing the observed radio flux density. We model the dynamics and emission of such an additional component following \citet{Panaitescu2000}, including the ejecta coasting phase, the trans-relativistic deceleration of the blastwave, the effect of synchrotron self-absorption on the emission (but still neglecting thermal electrons), and photon propagation effects. In Figure \ref{fig:narrow+wide}, we show the model light curve obtained by adding the flux densities predicted by such a model to those of the \citetalias{deWet2023} model, keeping the microphysical parameters fixed to values similar\footnote{\red{In \citetalias{deWet2023}, the best fit value of the slope of the relativistic electron energy distribution is $p = 2.0$. Since the model assumes that the electron energy distribution power law extends to infinity, this parameter choice is formally unphysical. Nevertheless, we find that by setting $p = 2.02$ we can reproduce the same results as \citetalias{deWet2023}.}}
to \citetalias{deWet2023}, that is, $\epsilon_\mathrm{e}=0.076$, $\epsilon_\mathrm{B}=4.6\times 10^{-3}$ and $p=2.02$. The wide jet component is assigned an isotropic-equivalent energy of $E_\mathrm{iso,w}=3\times 10^{53}$ erg, an initial bulk Lorentz factor of $\Gamma_{0,\mathrm
{w}}=3$, and a half-opening angle of $\theta_\mathrm{j,w}=15\degr$. Its collimation-corrected energy is therefore $E_\mathrm{w}= (1-\cos\theta_\mathrm{j,w})E_\mathrm{iso,w}\approx 10^{52}$ erg. For comparison, the narrow \citetalias{deWet2023} jet has $E_\mathrm{iso}=9\times 10^{53}$ erg and $\theta_\mathrm{j}=4.5\degr$ (and a formally infinite initial bulk Lorentz factor), hence a collimation-corrected energy of $E=(1-\cos\theta_\mathrm{j})\approx 3\times 10^{51}$ erg. The addition of the wide component therefore increases the energy budget by about a factor of four. 
The inclusion of this wider component leads to a clear improvement in the modelling of the late flattening in the radio light curves, as shown in the comparisons of the residuals in Figures~\ref{fig:220627a-lcsed}~and~\ref{fig:narrow+wide}. 

The presence of a slow, wide ejecta component that carries a similar (or even somewhat larger) collimation-corrected energy as the inner, narrow jet is expected in the case where the jet spends a large fraction of its total energy in breaking out of the progenitor star  \citep{Matzner2003,Bromberg2011,Harrison2018,Gottlieb2021}. 
\red{In such a scenario,} the energy is channelled to a hot cocoon that shrouds the jet during the propagation. After the jet breakout, the cocoon blows out producing a slow, wide ejecta component \citep{Salafia2022review}, which could explain the scenario behind our proposed modelling. For the cocoon to contain a large fraction of the total energy, the jet head must have propagated at a Newtonian velocity for most of its journey through the progenitor envelope \citep{Matzner2003,Bromberg2011,Salafia2022review,Gottlieb2021}. In such a regime, the time for the jet head to break out can be estimated based on the numerically-calibrated model of \citet[][their equation\ 16]{Harrison2018}, that gives a breakout time (relative to the birth time of the jet):
\begin{equation}
    t_\mathrm{bkt}\sim 37\,E_\mathrm{tot,52}^{-1/3}\,T_{3}^{1/3} \theta_\mathrm{base,10^\circ}^{4/3}\,R_{11}^{2/3}\,M_{10}^{1/3}\,\mathrm{s},
\end{equation}
where $E_\mathrm{tot,52}$ is the total energy in the ejecta (narrow plus wide) in units of $10^{52}$ erg, $T_{3}$ is the total jet duration in units of $1\,000$\,s, $\theta_\mathrm{base,10^\circ}$ is the half-opening angle of the jet at its base (i.e., at launch, before propagating through the stellar envelope) in units of $10\degr$, $R_{11}$ is the stellar radius in units of $10^{11}$\,cm, and $M_{10}$ is the stellar mass in units of $10\,\mathrm{M_\odot}$.

In this regime, the energy in the cocoon (and hence in the wide ejecta) is approximately 
$E_\mathrm{w}\sim E_\mathrm{tot}t_\mathrm{bkt}/T$. The total jet duration $T$ can be estimated as the sum of the breakout time plus the GRB duration, $T\sim t_\mathrm{bkt}+T_\mathrm{90}/(1+z)$. Using $E_\mathrm{tot}=E+E_\mathrm{w}$, these relations yield:
\begin{equation}
\begin{split}
& T \sim \frac{T_{90}}{1+z}\left(\frac{E_\mathrm{w}}{E}+1\right) \approx 1.2 \times 10^3\,\left(\frac{T_\mathrm{90}}{1100\,\mathrm{s}}\right)\left(\frac{1+z}{4.08}\right)^{-1}\\
& \indent\indent\indent\indent\indent\indent\indent\indent\indent\indent  
\times \left(\frac{E_\mathrm{w}/E+1}{4.3}\right)\,\mathrm{s},
\end{split}
\end{equation}
\begin{equation}
t_\mathrm{bkt} \sim \frac{T_{90}E_\mathrm{w}}{(1+z)E}\approx 9.0\times 10^{2}\,\left(\frac{T_\mathrm{90}}{1100\,\mathrm{s}}\right)\left(\frac{1+z}{4.08}\right)^{-1}\left(\frac{E_\mathrm{w}/E}{3.3}\right)\,\mathrm{s},    
\end{equation}
and
\begin{equation}
\begin{split}
& R \approx 1.3\times 10^{13}\,M_{10}^{-1/2}\left(\frac{T_\mathrm{90}}{1100\,\mathrm{s}}\right)\left(\frac{1+z}{4.08}\right)^{-1}\left(\frac{E_\mathrm{w}}{10^{52}\,\mathrm{erg}}\right)^{3/2}\\
& \indent\indent
\times\left(\frac{E}{3\times 10^{51}\,\mathrm{erg}}\right)^{-1}\theta_\mathrm{base,10^\circ}^{-2}\,\mathrm{cm}.  
\end{split}
\end{equation}
Here we used $E_\mathrm{w}$ and $E$ from the proposed modelling, $T_\mathrm{90}\sim 1100$\,s from \citetalias{deWet2023}, and we kept the initial half-opening angle at $10\degr$ (the jet is then expected to be collimated during propagation, see e.g.,\ \citealt{Bromberg2011}, \citealt{Salafia2022review}). 
\red{If indeed the wide component in our proposed modelling originates from a cocoon,} the long jet duration and large stellar radius obtained from these arguments \red{are then consistent with expectations for} 
the blue supergiant progenitor scenario proposed for ultra-long GRBs \citep[e.g.,][]{Gendre2013,Ioka2016,Perna2018}. 
In particular, we note that typical Wolf-Rayet stars have radii of only $\sim$1 R$_\odot$ 
compared to tens to a few hundred of R$_\odot$ for blue supergiants. 
This is similar to the conclusion reached by \citet{Piro2014} in their study of ultra-long GRB~130925A (with gamma-ray activity lasting for at least ${\sim} 7$\,ks); they found evidence for the presence of a hot cocoon which supported the blue supergiant progenitor scenario. 
\red{More recently, by conducting simulations of evolutionary pathways for different possible progenitor scenarios, the findings in \citet{Ror2024} similarly support blue supergiant stars as a plausible progenitor for ultra-long GRBs, including GRB~221009A which was the focus of their work.} 

\subsection{\fontsize{9.5}{12}\selectfont Testing the GRB~220627A Lensing Scenario with VLBI: Proof of Concept and Future Prospects}

The aim of our VLBA experiment was to provide a test of the lensing scenario for GRB~220627A that was independent of the conclusions drawn from the analysis of the high-energy spectra (see Subsection~\ref{ssec:220627a_fermigbm}, see also \citealt{Huang2022}). 
From the light curve evolution shown in Figure~\ref{fig:narrow+wide} (\textit{left}), we estimate the unresolved flux density at the time and frequency of our VLBA observation (15\,GHz, 11.7\,d post-burst) to be ${\sim} 150$\,\textmu Jy. 
This suggests that in the ultra-long GRB (non-lensing) scenario, we should expect a compact radio source at ${\sim} 150$\,\textmu Jy, and in the lensing scenario, two compact sources at even dimmer flux-density levels of ${\sim} 40$ and ${\sim} 110$\,\textmu Jy (as determined by their gamma-ray fluence brightness ratio $r~\approx~3$). 

Our VLBA observations yielded a non-detection down to a $5\sigma$ limit of \red{${\sim} 0.3$\,mJy\,beam$^{-1}$}. 
Within the uncertainty region of lower-resolution optical and radio detections (see Subsection~\ref{ssec:220627a_vlba}), there were a few noise peaks exceeding $3\sigma$ level, but none above the $5\sigma$ level. 
This means we were unable to claim a detection or a non-detection of an afterglow source (or any source) at the \red{${\sim} 0.3$\,mJy} level; the lensing and ultra-long GRB scenarios could not be distinguished so our VLBI experiment yielded a non-conclusive result.

This experiment was the first time VLBI was used to search for the multiple images of a  candidate strongly gravitationally lensed GRB afterglow. 
The only lensed GRB candidate that had VLBI follow-up prior to this event was GRB~200716C \citep{Giarratana2023}. 
For that event, as the afterglow had faded by the time of the observation, the European VLBI Network (EVN) and \textit{e}-Merlin were used to search the candidate host galaxy for a possible IMBH that could act as the lens, rather than directly for the multiple images of the afterglow. 
Their non-detection ($5\sigma$ upper limit of $< 50\,$\textmu Jy\,beam$^{-1}$) also yielded an inconclusive result, unable to confirm the presence of a possible IMBH lens in the candidate host galaxy. 
These results highlight the difficulty of such an experiment and the need for improvements in future programmes. 
Below, we evaluate the outcomes of our VLBI campaign for GRB~220627A and discuss the possible improvements that can be implemented in the areas of angular resolution, sensitivity, and trigger timing (including real-time correlation capabilities).

\begin{figure*}
\centering
\includegraphics[width=0.62\textwidth]{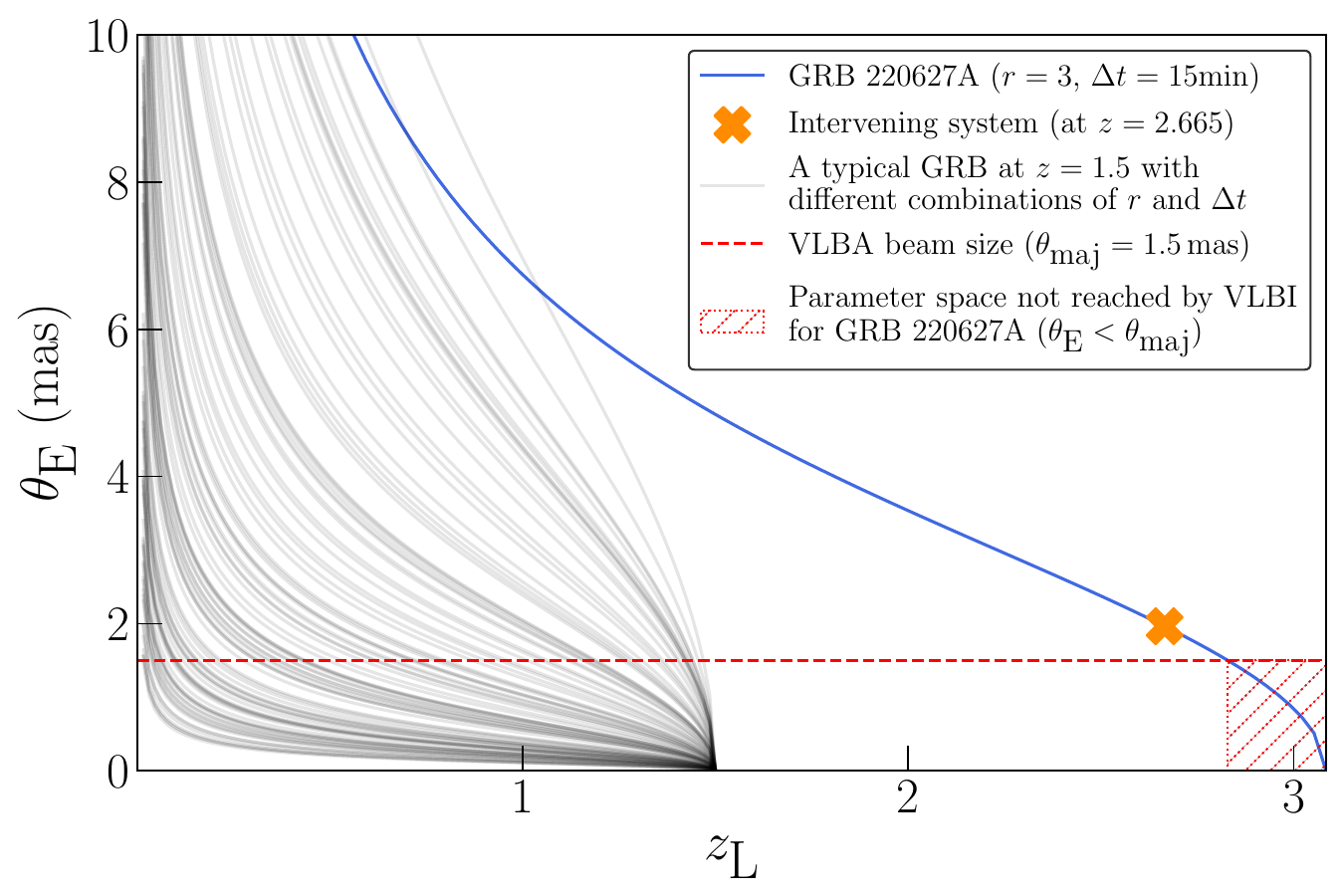}
\caption{
The Einstein radius $\theta_\textrm{E}$ as a function of the lens redshift $z_\textrm{L}$ for GRB~220627A is represented by the blue solid line. 
The expected Einstein radius for a lens located at the redshift $z=2.665$, which was the redshift of the intervening system identified in the optical spectrum \citep{Izzo2022GCN32291,deWet2023}, is plotted with an orange starred marker. 
These are compared against our VLBA major-axis beam size $\theta_\textrm{maj}$ represented here by the red dashed line. 
The part of the parameter space that cannot be probed by the angular limit of our observation is shown by the red hatched region. 
The Einstein radius as a function of lens redshift for a typical GRB at redshift $z=1.5$, given different, random (but possible) combinations of brightness ratios $r$ and time delays $\Delta t$ are also shown using grey solid lines. 
}
\label{fig:220627a_einstein-radius}
\end{figure*}

\subsubsection{Angular Resolution: probing different lens configurations}

Our resulting VLBA image had a synthesised beam size of $0.49\,\textrm{mas}~\times~1.51\,\textrm{mas}$. 
In Figure~\ref{fig:220627a_einstein-radius}, we show the expected Einstein radius for GRB~220627A in the lensing scenario as a function of the lens redshift, calculated using Equations~\ref{eq:time_delay} and \ref{eq:einstein_radius}, compared to our major-axis beam size. 
With the major-axis beam size being smaller than the expected Einstein radius out to $z \approx 2.8$ (n.b., $z_\textrm{GRB} = 3.084$), our VLBA observations had sufficient angular resolution to resolve the lensed images for much of the lens redshift parameter space. 
As an intervening absorber was identified at $z = 2.665$ via the presence of S \textsc{iv}, Si \textsc{ii}, C \textsc{iv}, Fe \textsc{ii}, Al \textsc{ii} lines in the optical spectrum \citep{Izzo2022GCN32291, deWet2023}, we considered this system the likely location of the lens during our radio campaign. 
At this redshift, the Einstein radius would be $\theta_\textrm{E} \approx 1.8$\,mas, so the two distinct lensed images of the afterglow (in the lensing scenario) would have been resolved by our VLBA observations. 
However, in the extreme scenarios, where the lens redshift is greater than $z_\textrm{L} \approx 2.8$, the lensed images may be unresolved. 
In this scenario -- if detected -- the lensed event could potentially still be confirmed by modelling a multi-component source, but this may be at the expense of accuracy in both the astrometry and source structure, which are critical for modelling of the lens (and by extension, time-delay cosmography). 

In Figure~\ref{fig:220627a_einstein-radius} we also plot 100 possible Einstein radii functions (i.e., as Einstein radius as a function of lens redshift) for a typical GRB at redshift $z = 1.5$ (the redshift where star formation and long GRB production is expected to be dominant). 
Each different Einstein radii function is produced by assuming different, random combinations of brightness ratio (selected from $1.1<r<5.0$) and time delay (selected from $0.6\,\textrm{s}\,< \Delta t <1\,000\,\textrm{s}\,$) for the $z=1.5$ GRB; the range chosen for these two variables reflect the observed range of values in previous claims of lensed GRB candidates. 
It is clear from this plot that the specific angular resolution required to test the lensing hypothesis for future candidates will be quite different depending on the lens configuration; nonetheless, for a large portion of such configurations, the milli-arcsecond scales probed by VLBI will be very useful for testing the lensing hypothesis. 

Future improvements to the programme should include the optionality to add longer baselines to the array, thereby improving the resolution; for example, adding EVN telescopes like Effelsberg to the VLBA to improve East-West $(u,v)$-coverage, or similarly, adding the East-Asian VLBI Array to the Long Baseline Array (LBA) in Australia to improve North-South $(u,v)$-coverage.

\subsubsection{Sensitivity: improvements with current and future arrays}

The more critical aspect for the GRB~220627A VLBI experiment is the sensitivity, as this was the limiting factor that led to the non-conclusive outcome. 
Two factors in particular need to be considered: the instrumentation and the brightness ratio in the scenario where the lensed images can be resolved. 
As shown in Subsection~\ref{ssec:radio-ag} (and in Figure~\ref{fig:narrow+wide}, \textit{right}), the radio spectrum below ${\sim} 15$\,GHz was self-absorbed. 
This meant that the LBA could not be triggered, 
despite its advantageous geographic position -- GRB~220627A is located at a Declination of $-32\degr$, while most LBA stations are located between a latitude of $-30\degr$ and $-40\degr$ -- 
because its optimal observing frequency is centred at $8.4$\,GHz, above which their receivers are significantly less sensitive. 
The VLBA was well-equipped to observe at 15\,GHz, but for our scenario, the observations were hindered by shorter integration times (with the source above the horizon for ${\sim} 6\,$h per day, c.f. ${\sim} 14\,$h per day for the LBA) and higher levels of tropospheric opacity ($< 30\degr$ elevation through the entire observation for most antennas due to the low Declination). 
The resulting image reached an rms sensitivity of $51$\,\textmu Jy\,beam$^{-1}$ (after combining two consecutive days of observations).
This sensitivity, as previously discussed, was insufficient to distinguish between the lensing and ultra-long GRB scenarios (as it does not reach the sensitivity level required to make a detection of the unresolved afterglow). 
With a lensing brightness ratio of $r \approx 3$, the sensitivity requirements in the lensing scenario for detecting both lensed images would be at least four times more severe than those required for only detecting the unresolved (and magnified) afterglow. 
With a more typical $r \approx 2$ from previous claims of GRB lensing candidates, the sensitivity requirements in the lensing scenario would then only be approximately three times more severe compared to the non-lensing scenario.

Our programme to confirm and study the properties of lensed GRB afterglows with VLBI serves as a pathfinder project for the SKA/ngVLA-era of radio astronomy. 
While the programme today is quite sensitivity limited, VLBI with the SKA is expected to reach rms sensitivities of ${\sim} 3\,$\textmu Jy\,beam$^{-1}$ in 1\,hr of integration for SKA1 and ${\sim} 0.05\,$\textmu Jy\,beam$^{-1}$ for SKA2 \citep[e.g.,][]{Paragi2015}. 
Similarly, VLBI with the ngVLA Large Baseline Array (ngVLA-LBA) is expected to reach rms sensitivities of ${\sim} 0.4\,$\textmu Jy\,beam$^{-1}$ in 1\,hr of integration \citep{Selina2018}. 
With this level of sensitivity, the SKA/ngVLA-era VLBI arrays should be able to make strong detections of the lensed images for the majority of GRBs regardless of their redshift, 
even if their afterglows are self-absorbed at the observing frequency \citep[see figure 2 in][]{Leung2021}. 
However, since the SKA and ngVLA telescopes will not be operating until later this decade at the earliest, our pathfinder programme needs other ways to improve instrumental sensitivity in the future triggers to successfully achieve its science goals. 
The main method for achieving this would be to ensure the trigger proposals have the option to add high-sensitivity telescopes to the array when the sensitivity requires it. 
For example, for GRB~220627A, if the observations had included the High Sensitivity Array (HSA), consisting of the Green Bank Telescope and the phased Very Large Array, our observations would have achieved an rms sensitivity\footnote{The HSA usually also includes Effelsberg, but this station does not operate at our Ku observing band. The estimated rms sensitivity reached by including the HSA was calculated here using the online tool maintained by Benito Marcote: \url{https://planobs.jive.eu/}} of ${\sim} 5\,$\textmu Jy\,beam$^{-1}$. 
This rms sensitivity would enable ${\sim} 8\sigma$ and ${\sim} 24\sigma$ detections of the two resolved lensed images in the lensing scenario, or a ${\sim} 32\sigma$ detection of an unresolved afterglow in the non-lensing, i.e., the ultra-long GRB afterglow, scenario (see Figure~\ref{fig:220627a-lcsed}, \textit{left}); this would have provided sufficient sensitivity for testing the lensing hypothesis. 

\subsubsection{Trigger Timing: the need for sensitive real-time capabilities}

\red{Triggering the telescope in a timely manner to catch the flux-density peak is important for maximising signal-to-noise of any detection.} 
If a GRB is able to be observed in a timely manner near the radio peak, the typical 10--30\,\textmu Jy rms sensitivity that the VLBA/LBA is able to achieve in a $10$\,hr observation without high-sensitivity telescopes in the array is sufficient for detecting the lensing signature for ${\sim} 75$\% of all expected events with a radio-loud counterpart \citep[i.e., those with integrated flux density reaches down to ${\sim} 200$\,\textmu Jy;][]{Chandra2012}. 

In the case of GRB~220627A, we planned to trigger the LBA a day after the ATCA observation if there was a detection at 9\,GHz, as the burst was optimally located in the Southern sky. 
However, the afterglow was not detected at 9\,GHz as the spectrum was self-absorbed and was instead detected at higher frequencies, which could be observed by the VLBA but not the LBA. 
Triggering the VLBA required more time and this led to a four-day delay between the ATCA observation and the high-resolution observation (as opposed to the one-day delay expected from triggering the LBA). 
The delay corresponded to a ${\sim} 200\,$\textmu Jy reduction in flux density, or a 53\% reduction in signal-to-noise. 
\red{This reduction in flux density was also steeper than what we expected due to incomplete multi-wavelength information at the beginning of the campaign, leading to a non-detection in our VLBA observations.}

Our GRB~220627A campaign therefore emphasises the requirement for future VLBI programmes to have a quick trigger turnaround time as well as the option to include high-sensitivity telescopes in the array, where possible (and needed). 
As the EVN and LBA are not full-time dedicated VLBI instruments like the VLBA, continuing to ensure sufficient operational flexibility (e.g., operating a subset of the array for VLBI in short notice) is important for these quick trigger turnarounds. We recognise both these observatories are making a strong effort on this front. 
Additionally, our campaign points to the increasing need for more sensitive e-VLBI (real-time correlation) capabilities, which would have allowed us to adjust our VLBI observing strategy \red{in a subsequent observation occurring even sooner} while the afterglow was still detectable by our instruments. 
Currently, the VLBA\footnote{\url{https://science.nrao.edu/facilities/vlba/docs/manuals/oss/referencemanual-all-pages}} is limited to a 128\,Mbps data recording rate for an e-VLBI setup, which does not provide sufficient sensitivity for extragalactic transient science, while the EVN\footnote{\url{https://www.evlbi.org/sites/default/files/shared/EVNCfP.pdf}} (LBA\footnote{\url{https://www.atnf.csiro.au/facilities/long-baseline-array/vlbi-national-facility-upgrade/}}) is capable of recording at 2\,Gbps (512\,Mbps) for an e-VLBI setup, but is instead limited by the range of frequencies it can observe at. 
Development of sensitive, responsive (in trigger turnaround and real-time correlation), and flexible (in frequency coverage) VLBI capabilities are therefore needed improvements for transient science, especially in the upcoming SKA/ngVLA era of radio astronomy. 

With event rates estimated to be as low as one every three years \citep{Oguri2019}, it is critical to ensure these improvements -- to angular resolution, sensitivity, and trigger timing (including real-time correlation capabilities) -- are made to maximise the science output of a gravitationally lensed GRB event. 
For example, one successful VLBI campaign for such an event would lead to the first confirmation of lensed GRB event (also the first milli-lens), provide a model for the lens, and also an estimate for the Hubble parameter, $H_0$. 
Assuming a typical event with the lensed images observed with an astrometric uncertainty of $\sim$100\,\textmu as, $H_0$ could be obtained to a $10\%$ precision \citep[e.g., equation 17 in][]{Birrer2019}; 20 to 30 such events (inclusive of other non-GRB explosive transients and variables) would enable a similar precision, and with different systematics, to other cosmology experiments \citep[e.g.,][]{Verde2019}. 

\section{Conclusions}
\label{sec:220627a_conclusions}

We conducted a radio campaign to study the afterglow of GRB~220627A, the most distant ultra-long GRB discovered to date, at $z=3.084$. 
The prompt gamma-ray light curve of GRB~220627A shows a double-pulse profile, with the pulses separated by a period of quiescence lasting approximately 15\,min, leading to initial speculation that GRB~220627A may be a gravitationally lensed GRB. 
Our comparison of the \textit{Fermi}/GBM spectra at time intervals corresponding to these pulses showed that their spectral properties differed by $> 5\sigma$ from each other; thus, we ruled out the lensing hypothesis and verified the ultra-long GRB classification (see also \citealt{Huang2022}). 

Our radio campaign led to the ATCA discovery of the radio afterglow at 7\,d post-burst. 
We continued to monitor it at multiple frequencies until 456\,d post-burst using the ATCA, MeerKAT, VLBA and VLA telescopes. 
\red{We found that there are three scenarios that could explain the observed radio properties, in particular the shallow radio decay ($F_{\nu} \propto t^{-1/2}$) following the much steeper decay ($F_{\nu} \propto t^{-2}$):
(i) energy injection from an additional, slower ejecta component (with Lorentz factor $\Gamma_\mathrm{sl-ej} \sim 10$) catching up to the external shock;
(ii) a stratified density profile going as $n \propto r^{-8/3}$; 
or alternatively, (iii) the presence of a slow, wide ejecta component (with collimation-corrected energy of $E_\mathrm{w} \approx 10^{52}$\,erg and a half-opening angle of $\theta_\mathrm{j,w}=15\degr$) in addition to a fast, narrow ejecta component (with collimation-corrected energy of $E\approx 3\times 10^{51}$\,erg and a half-opening angle of $\theta_\mathrm{j}=4.5\degr$)}.
\red{If we assume the wide component in scenario (iii) to originate from a cocoon,}
the cocoon properties would point towards a progenitor with a large radius ($R \approx 1.3 \times 10^{13} (M/10\,\mathrm{M_\odot})^{-1/2}\,\mathrm{cm}$), which 
\red{would be consistent with expectations for}
the blue supergiant preogenitor scenario proposed for ultra-long GRBs. 

While the lensing interpretation for this GRB was already ruled out after a detailed analysis of its high-energy spectra, we carried out VLBI observations as part of our radio campaign in an attempt to provide an independent verification of this conclusion in a way that was free from possible ambiguities from studying the light curves and/or spectra only. 
The results from our VLBI experiment yielded a non-detection (i.e., an inconclusive result), partly due to insufficient sensitivity and partly due to a fainter-than-expected afterglow brightness. 
The rms sensitivity of the experiment only reached $51$\,\textmu Jy\,beam$^{-1}$ at 15\,GHz partly due to the source's low elevation during the VLBA observations, while the decay of the radio flux density from ${\sim} 350$\,\textmu Jy, at 7\, post-burst, to ${\sim} 150$\,\textmu Jy, at 11\, post-burst, was steeper than anticipated (decaying as $F_{\nu} \propto t^{-2}$). 

Nonetheless, our experiment was the first time VLBI was used to search for the multiple images of a candidate strongly gravitationally lensed GRB afterglow. 
We have clearly demonstrated the need for improvements in several areas for similar VLBI experiments in the future, including but not limited to trigger latency, array flexibility, and real-time correlation capabilities. 
Implementing these improvements will ensure a successful VLBI confirmation of a lensed GRB event in the future, leading to possibly the first confirmation of a milli-lens and enabling an alternative method for estimating the Hubble parameter, $H_0$, with systematics that are different to other cosmology experiments. 

\section*{Acknowledgements}
We thank Paz Beniamini and also the anonymous referee for useful comments that improved the quality of this manuscript. 
We also thank Aaron Lawson and Tim Galvin for their useful advice on troubleshooting data problems with the VLA and ATCA observations, respectively. 
JKL, MRD, and AH acknowledge support from the University of Toronto and Hebrew University of Jerusalem through the University of Toronto -- Hebrew University of Jerusalem Research and Training Alliance program. 
MRD acknowledges support from the NSERC through grant RGPIN-2019-06186, the Canada Research Chairs Program, and the Dunlap Institute at the University of Toronto
The Dunlap Institute is funded through an endowment established by the David Dunlap family and the University of Toronto. 
GG and OSS acknowledge the European Union-Next Generation EU, PRIN 2022 RFF M4C21.1 (202298J7KT - PEACE). 
CS acknowledges financial support by the Istituto Nazionale di Astrofisica (INAF) Project 1.05.24.07.04 and by the Italian Ministry of University and Research (grant FIS2023-01611, CUP C53C25000300001).
TA acknowledges support from the Shanghai Oriental Talent Project,  Xinjiang Tianchi Talent Program and FAST special funding (NSFC 12041301). 
AH is grateful for the support by the the United States-Israel Binational Science Foundation (BSF grant 2020203) and by the Sir Zelman Cowen Universities Fund. This research was supported by the Israel Science Foundation (ISF grant No. 1679/23). 
Parts of this research were conducted by the Australian Research Council Centre of Excellence for Gravitational Wave Discovery (OzGrav), through project number CE230100016.
MPT acknowledges financial support from the Severo Ochoa grant CEX2021-001131-S and from the Spanish grant PID2023-147883NB-C21, funded by MCIU/AEI/ 10.13039/501100011033, as well as support through ERDF/EU. 

The Australia Telescope Compact Array is part of the Australia Telescope National Facility (\url{https://ror.org/05qajvd42}) which is funded by the Australian Government for operation as a National Facility managed by CSIRO.
We acknowledge the Gomeroi people as the Traditional Owners of the Observatory site.
The MeerKAT telescope is operated by the South African Radio Astronomy Observatory, which is a facility of the National Research Foundation, an agency of the Department of Science and Innovation.
The National Radio Astronomy Observatory is a facility of the National Science Foundation operated under cooperative agreement by Associated Universities, Inc.
This work made use of the Swinburne University of Technology software correlator, developed as part of the Australian Major National Research Facilities Programme and operated under licence.
This work made use of data supplied by the UK Swift Science Data Centre at the University of Leicester.

\section*{Data Availability}
The ATCA data can be accessed through the Australia Telescope Online Archive\footnote{\url{https://atoa.atnf.csiro.au}} under the project code C3478.
The MeerKAT data can be accessed through the MeerKAT data archive\footnote{\url{https://archive.sarao.ac.za}} under project code DDT-20220705-SG-01. 
The VLA and VLBA data can be accessed through the NRAO Archive Interface\footnote{\url{https://data.nrao.edu}} under project codes VLA/22B-031 and VLBA/22B-032, respectively.
Reasonable requests for other auxiliary datasets can be accommodated for via email to the corresponding author.

\vspace{5mm}
\facilities{ATCA, \textit{Fermi}/GBM, MeerKAT, \textit{Swift}/XRT, VLA, VLBA}

\bibliography{references}{}
\bibliographystyle{aasjournal}

\end{document}